\documentclass[]{interact}

\usepackage{algorithm}%
\usepackage{algorithmicx}%
\usepackage{algpseudocode}%
\usepackage{setspace}%
\usepackage{epstopdf}
\usepackage{graphicx}
\usepackage{afterpage}
\usepackage{rotating} 
\usepackage{subfigure}
\usepackage{hyperref}
\usepackage{cleveref}
\hypersetup{
    colorlinks=true,
    linkcolor=blue,
    filecolor=magenta,      
    urlcolor=cyan,
    pdftitle={Overleaf Example},
    pdfpagemode=FullScreen,
    }
\usepackage{wrapfig}
\usepackage{amsmath}
\usepackage{amssymb}
\usepackage{mathtools}

\newcommand{\sm}[1]{{\scriptscriptstyle#1}}
\usepackage{accents}

\MakeRobust{\underaccent}

\newcommand{\lt}[1]{\underaccent{\tilde}{#1}}
\newcommand{\LT}[1]{\lt{#1}}

\newcommand{\ut}[1]{\tilde{#1}}
\newcommand{\UT}[1]{\ut{#1}}

\newcommand*{\myfont}{\fontfamily{pcr}\selectfont}

\usepackage{natbib}
\bibpunct[, ]{(}{)}{;}{a}{}{,}
\renewcommand\bibfont{\fontsize{10}{12}\selectfont}

\theoremstyle{plain}

\theoremstyle{definition}

\theoremstyle{remark}

\usepackage{booktabs}
\usepackage{hhline}
\usepackage{multirow}

\begin{document}


\title{Confidence Intervals on Multivariate Normal Quantiles for Environmental Specification Development in Multi-axis Shock and Vibration Testing}

\author{
\name{Adam Watts\textsuperscript{a}\thanks{CONTACT Adam Watts. Email: acwatts@lanl.gov}, Thomas Thompson\textsuperscript{a}, and Dustin Harvey\textsuperscript{a}}
\affil{\textsuperscript{a}E-14 Test Engineering, Los Alamos National Laboratory, P.O. Box 1663, Los Alamos,
87545, New Mexico, USA.}
}

\maketitle

\begin{abstract}
This article describes two Monte Carlo methods for calculating confidence intervals on cumulative density function (CDF) based multivariate normal quantiles that allows for controlling the tail regions of a multivariate distribution where one is most concerned about extreme responses. The CDF based multivariate normal quantiles associated with bivariate distributions are represented as contours and for trivariate distributions represented as iso-surfaces. We first provide a novel methodology for an inverse problem, characterizing the uncertainty on the $\tau^{\mathrm{th}}$ multivariate quantile probability, when using concurrent univariate quantile probabilities. The uncertainty on the $\tau^{\mathrm{th}}$ multivariate quantile probability demonstrates inadequacy in univariate methods which neglect correlation between multiple variates. Limitations of traditional multivariate normal tolerance regions and simultaneous univariate tolerance methods are discussed thereby necessitating the need for confidence intervals on CDF based multivariate normal quantiles. Two Monte Carlo methods are discussed; the first calculates the CDF over a tessellated domain followed by taking a bootstrap confidence interval over the tessellated CDF. The CDF based multivariate quantiles are then estimated from the CDF confidence intervals. For the second method, only the point associated with highest probability density along the CDF based quantile is calculated, which greatly improves the computational speed compared to the first method. Monte Carlo simulation studies are used to assess the performance of the various methods. Finally, real data analysis is performed to illustrate a workflow for CDF based multivariate normal quantiles in the domain of mechanical shock and vibration  to specify a minimum conservative test level for environmental specification.

\end{abstract}

\begin{keywords}
Quantile; Tolerance; Multivariate; Coverage; Multi-axis; Specification
\end{keywords}

\section{Introduction}\label{Intro}
Statistical quality control has many facets, one of which are defining the limits or bounds that a manufactured product must adhere to in order for that product to meet quality criteria. In the domain of mechanical shock and vibration, deriving the upper limit on the vibration or shock environmental levels that could potentially result in damage to a system or component are of concern. In single-axis testing, the test item is exposed to a single-axis excitation and then is rotated and re-mounted to test the subsequent orthogonal axes. Traditional shock and vibration testing only considers the univariate case for a single-axis response variable. Multi-axis testing is becoming more prominent as it provides excitation from more than one axes and doesn't require for the test body to be rotated and re-mounted. Tolerance intervals have been utilized to specify a conservative upper bound as the minimum testing levels for single-axis environmental tests as they can be estimated from using sample sizes as low as $n=3$, but do not take into account correlation between variates as they are a univariate statistic method. The methodology presented here offers a new statistical method for deriving these limits when working with multiple variates which are either response locations and/or response directions while including the correlations between all the variates. 

\begin{figure}[h]
\includegraphics[width=8cm]{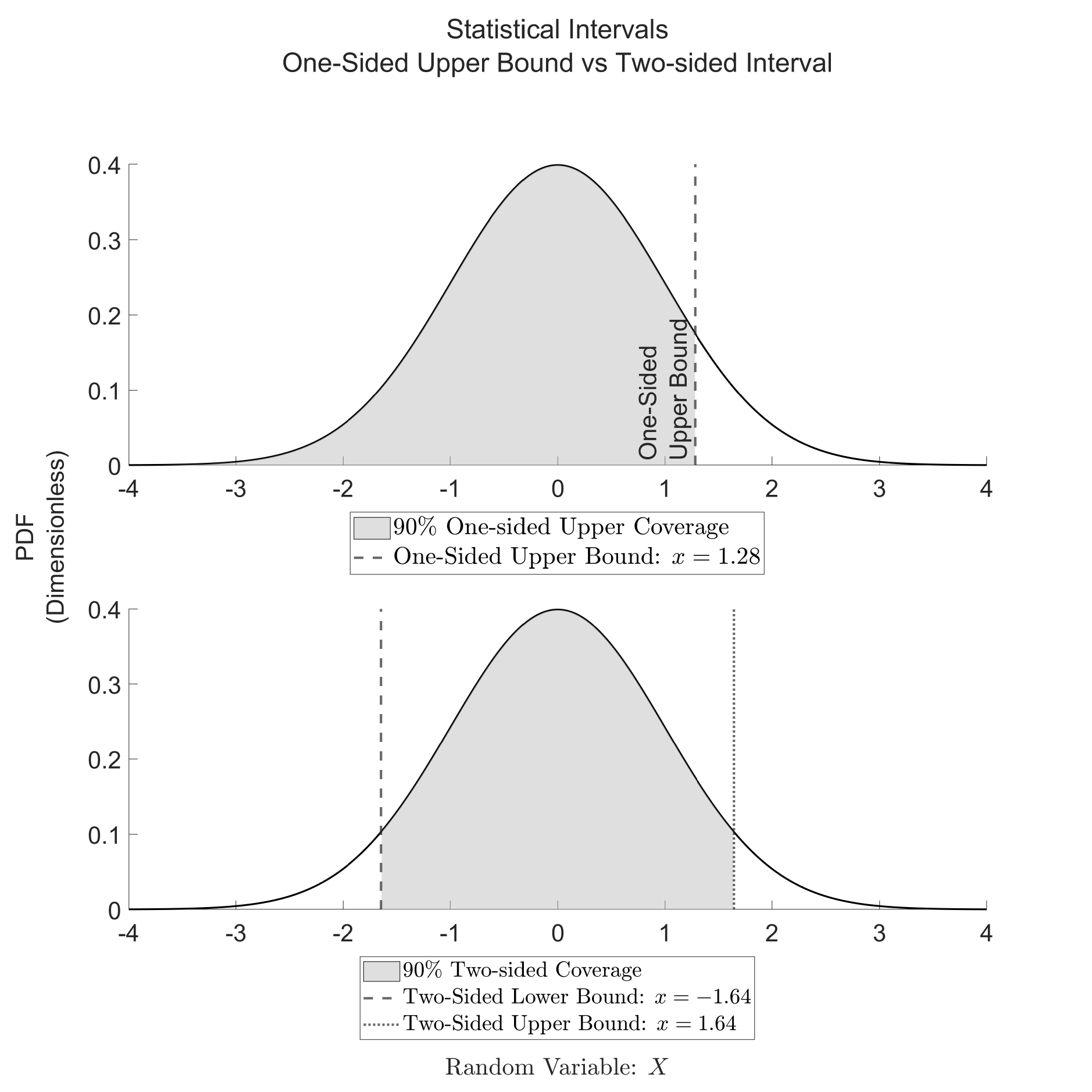}
\centering
\caption{Comparing a two-sided and one-sided statistical interval on the bounds covering 90\% of the population using the standard Normal distribution. As the sample size approaches infinity, $n \rightarrow \infty$, the one-sided tolerance bounds for a coverage value of $\beta=0.9$ and the one-sided confidence interval for the quantile probability of $\tau=0.90$ both converge to a value of $x=1.28$, see \eqref{EQ:TOL_UPPER} and \eqref{EQ:CI_QUANTILE_UPPER}.}
 \label{Plot:OneSided_vs_TwoSided}
\end{figure}

Traditional univariate tolerance intervals were first introduced in the early 1940s by the statistician Abraham Wald\citep{wald_tolerance_1946, wald_extension_1943, wald_setting_1942}. A normal tolerance interval is a statistical interval that contains at least a proportion, $\beta$, of the distribution or population for a given confidence, $\gamma$, where $\gamma=1-\alpha$ and $\alpha$ is the significance level used to compute the confidence level. The term \textit{normal} denotes that the data can be represented as originating from a population that is normally distributed. We will drop the normal prefix from tolerance interval and tolerance bound for brevity. One-sided tolerance bounds are equivalent to a one-sided confidence interval for a normal distribution quantile \citep{meeker_statistical_2023}. The easiest example of this equivalency occurs for an upper statistical limit. In this case, the one-sided upper tolerance bound for the proportion of the population, $\beta$, and the one-sided confidence bound on the $\tau^{\mathrm{th}}$ quantile probability, both imply encompassing the population starting at $-\infty$ and ending at the same value. Figure \ref{Plot:OneSided_vs_TwoSided} demonstrates the difference between a one-sided statistical bound and two-sided statistical interval covering 90\% of the population. One should note how the one-sided upper bound includes the lower tail of the distribution; a one-sided lower bound includes the upper tail of the distribution. When the distribution’s parameters are known, a one-sided tolerance bound, or equivalently, a one-sided confidence bound on the distribution's quantile associated with the $\tau^{\mathrm{th}}$ quantile probability, is just the inverse cumulative distribution function (ICDF). The ICDF is also known as the quantile function, denoted as $Q(\bullet)$. A tolerance interval or bounds accounts for the uncertainty in the parameters of the population's distribution as it incorporates the sample size, $n$, directly into its calculation. As the sample size approaches infinity, $n \rightarrow \infty$, a one-sided tolerance bound converges to the ICDF, regardless of the desired confidence due to having zero uncertainty on the parameters of the population.

Recent simulation work from \cite{chiang_tolerance_2021} have confirmed the theoretical properties of tolerance intervals. Univariate tolerance intervals can be found among the industrial and medical domains \citep{kenett_modern_2021, francq_tolerate_2020}. There still exists much confusion over the variety of statistical intervals as many engineers are familiar with confidence intervals only in the context of the distribution mean. However, confidence intervals can be applied to other parameter or characteristic of a distribution that represents a population or process. For example a confidence interval for the distribution median, standard deviation, or quantile. In other contexts, an enclosure interval such as a two-side tolerance interval that contain, or envelopes, at least a specified proportion of a distribution is required. Perhaps some of the mistakes around selecting the correct statistical interval is because, ``... statistics textbooks typically focus on the common confidence intervals, occasionally make reference to tolerance intervals, and consider prediction intervals only in the context of regression analysis. This is unfortunate because in applications, tolerance intervals, prediction intervals, and confidence intervals on distribution quantiles and on exceedance probabilities are needed almost as frequently as the better-known confidence intervals." \citep{meeker_statistical_2023}. 

Two-sided statistical intervals have both a finite lower endpoint and finite upper endpoint. Each endpoint has an equal degree of uncertainty associated with the parameter, or quantity of interest, located at each end of the interval's endpoints \citep{meeker_statistical_2023}. One-sided statistical bounds on the other hand have a degree of uncertainty only associated with either the lower endpoint for a one-sided lower bound or the upper endpoint for a one-sided upper bound. One-sided statistical bounds do not have finite endpoints with a one-sided lower bound having an upper endpoint of $+\infty$ and a one-sided upper bound having a lower endpoint of $-\infty$. One can approximately make a one-sided lower and one-sided upper confidence interval from a two-sided confidence interval. For example, a 90\% two-sided confidence interval is approximately equivalent to a 95\%, one-sided, upper and lower confidence interval respectively \citep{meeker_statistical_2023}. 

For univariate normal distributions, the quantile, is the value associated to the quantile function on the $\tau^{\mathrm{th}}$ quantile probability. In univariate statistics, the quantiles are finite scalars, $ Q(\tau) \rightarrow (- \infty, \infty ): 0 < \tau  < 1$. In the multivariate case, quantiles, are the set of vectors from the quantile function for the $\tau^{\mathrm{th}}$ quantile probability, $Q(\tau)$. The quantiles in the multivariate case are now an infinite set of vectors, $\boldsymbol{x}$:
\begin{align}
    Q(\tau) =\left\{ \boldsymbol{x} = (x_1, \ldots, x_q) \in \mathbb{R}^{q} \ : \  \tau = F(\boldsymbol{x})   \right\}, \label{EQ:MV_Quantile}
\end{align}
where $ F(\boldsymbol{x})$ is the CDF function for a $q$-variate normal distribution. The term \textit{univariate quantile} refer to the value associated to the quantile function on the $\tau^{\mathrm{th}}$ quantile probability, i.e., $x=Q(\tau)$. The term \textit{CDF based multivariate quantile}, or just \textit{CDF based quantile}, refer to the infinite set of vectors associated to the quantile function for the $\tau^{\mathrm{th}}$ quantile probability, \eqref{EQ:MV_Quantile}.

One-sided (upper/lower) and two-sided statistical bounds are unambiguous in univariate statistical analysis but not in the multivariate case. We shall see in section \ref{Elliptical_Tolerance}, traditional normal tolerance regions bound the population data from the distribution's center (mean) outwards \citep{lucagbo_rectangular_2023, young_nonparametric_2020}. These multivariate regions can be thought analogously as a multivariate extension of the ``two-sided'' tolerance interval utilized in univariate statistics. Nonetheless, there does not appear to be a multivariate extension of the ``one-sided'' tolerance bound or equivalently a methodology for constructing confidence intervals on CDF based multivariate quantiles which would allow for controlling the tail regions of a multivariate normal distributions. It is at these tail regions where one is most concerned about controlling extreme responses for a given confidence value.

Tolerance regions associated with multivariate normal distributions are thoroughly described by \cite{chew_confidence_1966} with the most recent and canonical being from \cite{krishnamoorthy_multivariate_2009}. \cite{bornkamp_calculating_2018} currently provides one of the few methodologies for calculating CDF based multivariate quantiles. The caveat with the method is that it only produces a single, equicoordinate vector, within the infinite set of vectors that comprise a multivariate quantile. When used appropriately by standardizing a multivariate distribution, the equicoordinate vector can be used very efficiently. \cite{chen_distribution-function-based_2002} provide methodologies for multivariate quantiles mostly applicable for the bivariate case but provides little in how to characterize uncertainty on the generated quantiles. \cite{klein_directional_2020} provide a detailed methodology for CDF based bivariate quantiles by transforming the bivariate data to the unit square but also don't address uncertainty. 

The purpose and scope of this paper is extended over several sections. \Cref{Inadequacy} demonstrates the inadequacy of using concurrent univariate tolerance intervals or concurrent univariate quantiles in a multivariate space. We present a novel methodology for characterizing the uncertainty on the $\tau^{\mathrm{th}}$ multivariate quantile probability when utilizing concurrent univariate tolerance intervals or univariate quantiles. \Cref{Tolerance_Regions} presents classical tolerance regions and introduces the concept of CDF based multivariate quantiles as either contours, iso-surfaces, or hyper iso surfaces. \Cref{CI_MV_Quantiles} discusses the importance of standardization in multivariate statistics. We demonstrate how to construct confidence intervals on CDF based multivariate normal quantiles with an algorithm for the trivariate case. The concept of a critical point, the point residing on a multivariate quantile that has the highest relative likelihood (maximum PDF value) of occurring, is introduced. Another algorithm is provided for constructing confidence intervals for just the critical points on CDF based multivariate quantiles. Confidence intervals on a critical point greatly speeds up the computation time compared to the previous method which estimates portions of the CDF based multivariate quantile. Monte Carlo simulation studies and a case study in the field of mechanical shock is carried out in \cref{Results}. Finally, we describe further research directions summarize this paper in \cref{Conclusion}. 

Since multivariate normal distributions are the ubiquitous multivariate distribution, we believe that developing confidence intervals on CDF based multivariate normal quantiles and confidence intervals for critical points on these multivariate quantiles greatly benefits a large audience, particularity those working in field of quality control.

\section{Inadequacy of Univariate Methods in a Multivariate Space}\label{Inadequacy}

Univariate, one-sided, upper tolerance bounds are commonly used in quality control. The tolerance bound is denoted below, 
\begin{align}
    \UT{T}_{\beta} (1-\alpha; \beta, n) \ ,\label{EQ:TOL_UPPER}
\end{align}
where $1-\alpha$ is the confidence value, $\beta$ is the proportion of the population, and $n$ denotes the sample size. When the tilde is underneath the variable such as, $\LT{T}$, it denotes a one-sided lower bound. When the  tilde is above the variable it denotes a one-sided, upper, bound, such as $\UT{T}$. Equation \eqref{EQ:TOL_UPPER} is equivalent to a one-sided, upper, confidence bound for the $\tau^\mathrm{th}$ quantile probability which is denoted below 
\begin{align}
    \UT{x}_{\tau} (1-\alpha; \tau, n) \ . \label{EQ:CI_QUANTILE_UPPER}
\end{align}

For the example provided below, each variate corresponds to one of the three spatial directions: X, Y, and Z. One might be inclined to use multiple, one-sided, confidence bounds for the $\tau^\mathrm{th}$ quantile, $Q(\tau_{i}) : i \in \left\{X, Y, Z\right\}$, when working with multiple variates. Using multiple, one-sided bounds, for each spatial direction, leads to uncertainty as to what is the exact multivariate quantile probability, $\tau_{\sm{XYZ}}$, over the entire space because the correlations between the variates are neglected. Looking at this predicament through the lenses of a multiple hypothesis problem, one can bound the multivariate quantile probability for all three variates.

For example, if $\tau_{i} = 0.90 : i \in \left\{X, Y, Z\right\}$, then the lower and upper bounds on the multivariate quantile probability are, $ 0.70 \leq \tau_{\sm{XYZ}} \leq 0.90$. The uncorrelated case for this example would have a multivariate quantile probability of $\tau_{\sm{XYZ}} = (0.90)^3 = 0.729$, which is between the bounds shown above. See \cref{appendix:a} for the equations to calculate these probabilities.

\begin{figure}[H] 
\includegraphics[width=12cm]
{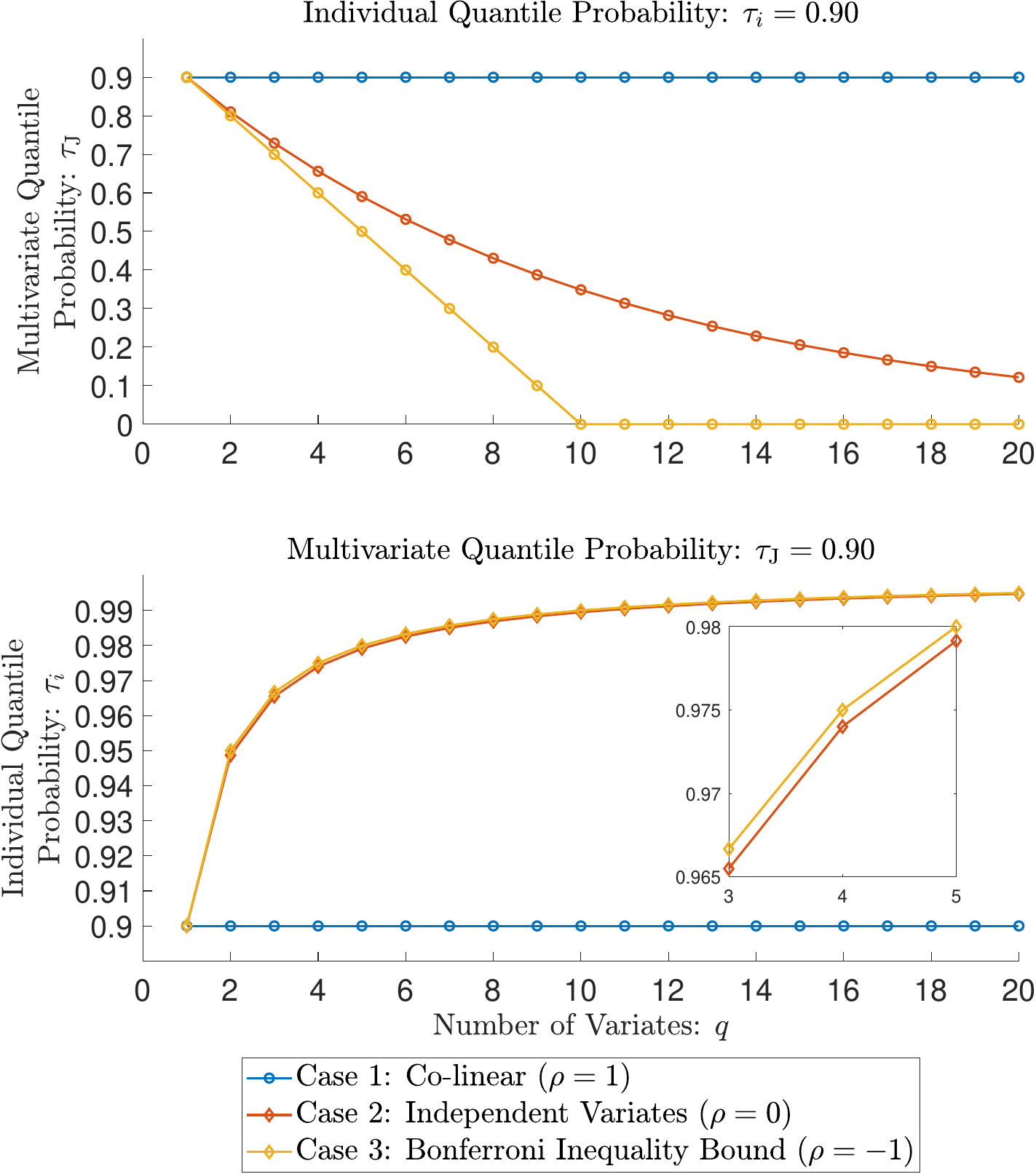}
\centering
\caption{Top plot shows how the multivariate or joint quantile probability, $\tau_{\mathrm{J}}$, decreases assuming a constant individual quantile probability value, $\tau_{i} = 0.90$. At ten variates, $q=10$, the Bonferroni inequality on the lower bound on the joint quantile probability reaches zero, $\tau_{\mathrm{J}}=0$. The Bonferroni inequality bound occurs when the correlation among all the variates are maximally negative. However, when all the variates are independent, the joint quantile probability only equals zero at the limit as the number of variates reaches infinity, $ \lim\limits_{q \to \infty} (0.90)^q = 0$. Case 1 is the trivial case when the variates are maximally, positively, correlated ($\rho=1$). The bottom plot shows what the individual quantile probability values must be adjusted to for both cases in order to achieve a constant joint quantile probability value, $\tau_{\mathrm{J}}=0.90$. The inset plot, found in the bottom plot, demonstrates that the individual quantile probability values must be adjusted slightly \textit{more} for case 3 than for case 2.}
\label{Plot:MultipleHypothesis}
\end{figure}

The top plot of Figure \ref{Plot:MultipleHypothesis} depicts how the multivariate quantile probability decreases as the number of variates increases from 1 to 20. One of the most common ways to alleviate the drop in the multivariate quantile  probability is to adjust the individual quantile probability values based on either the Bonferroni inequality or assuming that the variates are independent, thus uncorrelated. The bottom plot of Figure \ref{Plot:MultipleHypothesis} depicts how the individual quantile probability values must increase as the number of variates increases from 1 to 20 such that the multivariate quantile probability maintains a constant value, $\tau_{\mathrm{J}}=0.90$. One can see how the individual quantile probability values, $\tau_{i}$, must increase from their original quantile probability value in order to maintain a constant multivariate quantile probability. When the variates are independent (case 2) the individual quantile probabilities need to be adjusted less than when the variates are maximally, negatively, correlated using Bonferroni inequality (case 3). Therefore, the individual quantile probability values must be adjusted slightly \textit{more} for case 3 than for case 2, since the probability of the intersection of independent events is the product of their probabilities. Case 1 is the trivial case when the variates are maximally, positively, correlated ($\rho=1$).

\subsection{Calculating the Joint Quantile Probability from a Known Distribution}\label{The_Effects_of_Correlation}
Instead of bounding the quantile probability for the joint distribution, one can numerically calculate the exact quantile probability when the variates can be modeled from a known multivariate normal distribution. This can be done when using concurrent univariate quantiles. The term \textit{concurrent} implies that each variate utilizes the same quantile probability, i.e., $\tau_x = \tau_y = \tau_z$. The numerical method works by utilizing equicoordinate quantiles. To demonstrate this, an example in $\mathbb{R}^2$ will be used with the standard normal bivariate distribution,
\begin{align}
\begin{pmatrix} X \\ Y \\  \end{pmatrix}  \sim   \mathcal{N} \left[ \boldsymbol{\mu} = \begin{pmatrix} 0 \\ 0  \end{pmatrix},   \Sigma = \begin{pmatrix} 1 & \rho \\ \rho & 1 \end{pmatrix}    \right],
\end{align}
where $\rho$ denotes the correlation between the two random variables, $X$ and $Y$. In the example provided, the marginal distributions for $X$ and $Y$ are the just the standard normal distribution, $X \sim \mathcal{N}(0, 1)$ and $Y \sim \mathcal{N}(0, 1)$. Since the parameters associated for each marginal distribution are known, the one-sided upper tolerance bound is equivalent to the $\tau^\mathrm{th}$ quantile probability, $Q(\tau=0.90) = 1.2816$. Using equations \eqref{EQ:Coliner} and \eqref{EQ:Bonferroni} while assuming an individual quantile probability value of $0.90 = \tau_{i} : i \in \left\{X, Y\right\}$, implies that the lower and upper bounds for the joint quantile probability are, $ 0.80 \leq \tau_{\sm{XY}} \leq 0.90$.

\begin{sidewaysfigure}[h]
\centering
\includegraphics[width=\textwidth, keepaspectratio]{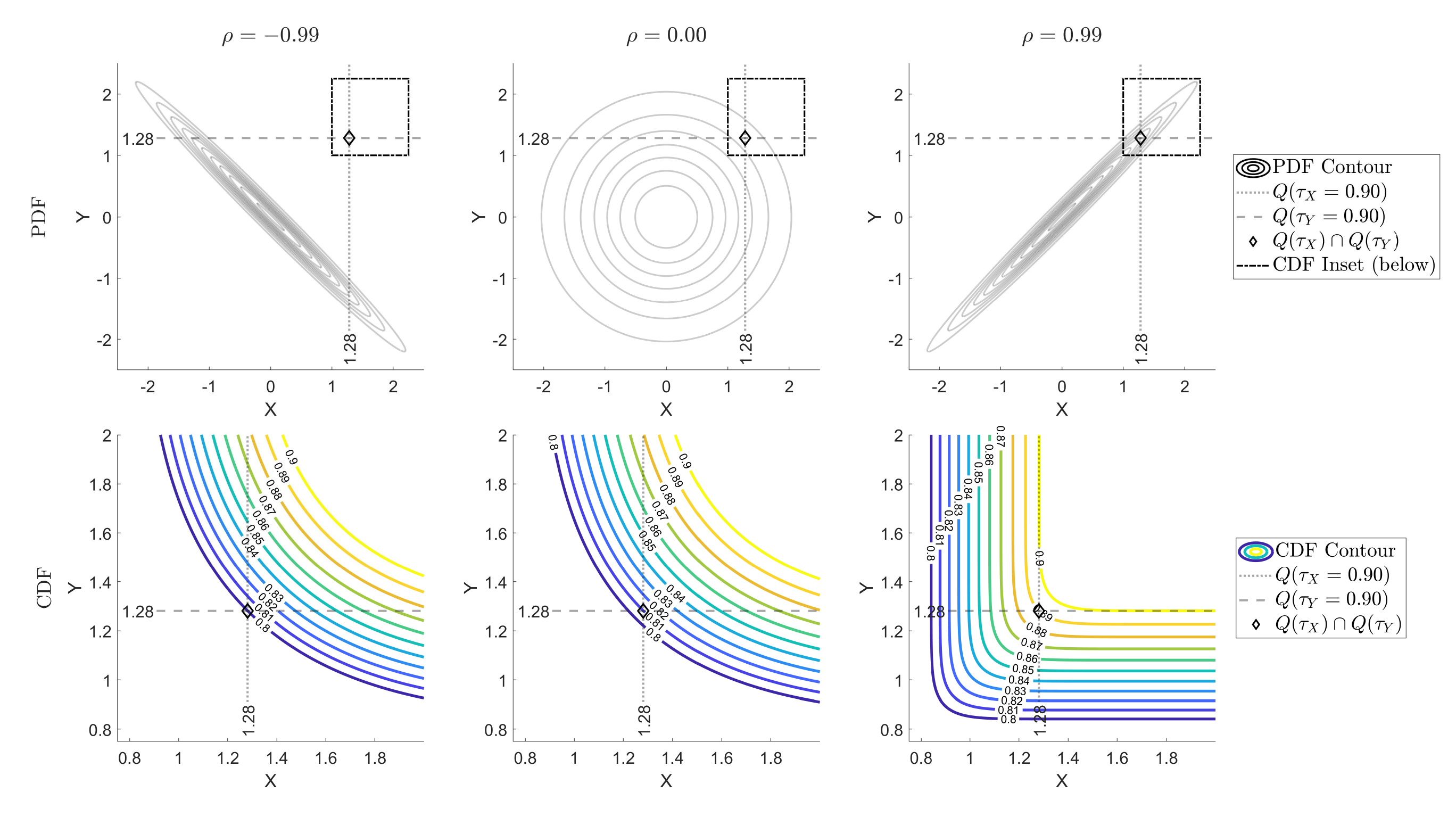}
\centering
\caption{The first row depicts the PDF contours, shown in grey, of three standard normal bivariate normal distributions with correlation values of $\rho \in \left\{-0.99, 0, 0.99 \right\}$, one for each column. The black dash-dotted line are the axis limits for the inset CDF contour colors below. The second row depicts the CDF contours, shown in color, for each standard normal bivariate distributions with different correlations. The dotted and dashed lines correspond to the univariate ICDF, $Q(\tau_{i})$, at a quantile probability of $\tau_{i} = 0.90 : i \in \left\{X , Y\right\} $. The diamond represents the intersection of the two univariate ICDF lines.}\label{Plot:CorrelationContour}
\end{sidewaysfigure}

In the case where the variates are uncorrelated, $\tau_{\sm{XY}} = 0.90^2 = 0.81$. The solution can be graphically seen in the second row, middle column, of Figure \ref{Plot:CorrelationContour}. One can visually see that the intersection of each univariate quantile lines and the CDF contour line occurs at the multivariate quantile probability at a value of 0.81. For the case when $\rho=-0.99$ and $\rho=0.99$, the intersection of each univariate quantile line and the CDF contour lines occurs at the multivariate quantile probability at a value of 0.800 and 0.8901 respectively. The solutions are graphically shown in the second row, first and third column, respectively, of Figure \ref{Plot:CorrelationContour}. 

Numerical solutions can be had using the {\myfont mvtnorm} library in the R language which provides a function called {\myfont qmvnorm()} which  computes the equicoordinate quantile function of the multivariate normal distribution for arbitrary correlation matrices based on inversion of the probability distribution function for the multivariate normal distribution. The algorithm utilizes a stochastic root finding algorithm described in \cite{bornkamp_calculating_2018}. The numerical solution can only be computed such that all the variates share the same quantile probability, hence, equicoordinate, i.e.; $\tau = \tau _{i} : i \in \left\{1, \ldots ,q \right\}\ $ for $q$ number of variates. If the above condition holds, then the numerical solution for the joint quantile probability for arbitrary normal multivariate distributions can be written as,
\begin{gather}
g(p; \tau, C_{\sm{XY}}) = \bigl(Q(\tau) - \text{{\myfont qmvnorm}}(p, \mathrm{corr}=C_{\sm{XY}}, \mathrm{tail = ``lower.tail"}) \bigl)^2,  \label{EQ:MinFunct} \\ 
G(p) = \underset{p}{\arg\min} \biggl[ g(p; \tau, C_{\sm{XY}}) \biggl] ,  \label{EQ:Min}
\end{gather}
where $C_{\sm{XY}}$ is the correlation matrix, \textit{tail}, corresponds to the optional input argument for the function, and $p : 0 \leq x \leq 1$, is the space over all quantile probability values associated to the multivariate CDF. Since {\myfont qmvnorm()} computes the equicoordinate quantile value, this implies that the function returns just a scalar. For example, in the bivariate case where $\rho=0$, the joint quantile probability value of $p$ that minimizes \eqref{EQ:Min} is $p=0.81$. Therefore the equicoordinate quantile value for this bivariate distribution would be the X, Y coordinate pair $(1.2816, 1.2816)$ as shown by the function output below,
\begin{align}\
 1.2816 = \text{{\myfont qmvnorm}}(0.81, \mathrm{corr}= \begin{pmatrix} 1 & \rho \\ \rho & 1 \end{pmatrix}, \mathrm{tail = ``lower.tail"}). \nonumber
\end{align}
 When $\rho=0$, this is one of the few instances where one can analytically solve for the joint quantile probability, $\tau_{\sm{XY}}=0.81$, without needing numerical methods.

\afterpage{\clearpage}

\subsection{Characterizing Uncertainty on the Joint Quantile Probability}\label{UQ_on_their_Joint_Quantile}
The true covariance or correlation matrix is rarely known for data that is multivariate normally distributed. Since there is uncertainty on the true correlation matrix, there will also be uncertainty on the multivariate quantile probability. We present here a methodology for characterizing the uncertainty on the multivariate quantile probability by comparing to concurrent univariate quantile probabilities, Algorithm \ref{ALGO:UQ_JointQuantile}. The methodology utilizes a nonparametric bootstrap to estimate uncertainty by estimating the sampling distribution on the multivariate quantile probability \citep{efron_bootstrap_1979, efron_bootstrap_1986}.

For each nonparametric bootstrap sample, the data is standardized. Standardization implies that for each bootstrap sample, the bootstrap data is centered such that its mean is the zero vector, $\boldsymbol{\mu}=\boldsymbol{0}$. It also implies that the bootstrap data is scaled such that the variance for each variate is unity, i.e., $\mathrm{diag}(S^*) = \boldsymbol{1}_{q}$, where $S^*$ denotes the bootstrap sample covariance. The standardization process implies that the bootstrap sample correlation matrix equals the bootstrap sample covariance matrix, $C^* = S^*$. Standardizing the bootstrap data allows one to utilize the quantile function for the standard normal distribution for the individual variates to calculate the random vector, $x$, such that its probability is less than or equal to $\tau$. The standardization process alleviates the need for characterizing uncertainty on the mean and variance, thus allowing to exclusively characterize the affects of correlation on the multivariate quantile probability. Algorithm \ref{ALGO:UQ_JointQuantile} is thus simply resampling many possible correlation matrices from the original data in a transformed space that allows one to exploit the quantile function for the standard normal distribution.

In \cref{Inadequacy} we have shown how to calculate uncertainty on the upper and lower bound for the multivariate quantile probability when one neglects correlations between the variates. Algorithm \ref{ALGO:UQ_JointQuantile} provides a new methodology, thereby improving the characterization of the uncertainty on the multivariate quantile probability when the population is assumed to originate from a multivariate distribution. If practitioners are going to use concurrent univariate tolerance or concurrent univariate confidence intervals for the distribution's quantile when working with multivariate data, it is highly advised to use algorithm \ref{ALGO:UQ_JointQuantile} to assess the uncertainty on the multivariate quantile probability and whether that level of uncertainty is acceptable. Ideally, with the methods presented in \cref{Method1} and \cref{Method2} one should utilize the proper multivariate statistical interval when working with multivariate data and not revert to single variate statistical intervals.

\begin{algorithm}[H]
\caption{ - R pseudo code for calculating the uncertainty on the joint multivariate quantile probability, $\tau_{\gamma}$, by comparing to concurrent univariate quantile probabilities, $\tau_{i}$. $\gamma$ are the $\gamma^\mathrm{th}$ percentile values used to construct the simple percentile bootstrap confidence intervals on the joint quantile probability, $\tau_{\gamma}$. $D \in \mathbb{R}^{ (n \times q)}$ is a matrix of data originating from a $q$-dimensional multivariate distribution with $n$ samples.  Since the data is scaled with unity variance along the diagonal, this implies the correlation matrix equals the sample covariance matrix, $C^* = S^*$.}
\label{ALGO:UQ_JointQuantile}
    \begin{algorithmic}[1]    
    \Require $0 < \tau_i < 1, \quad  0 < \gamma < 1$
    \Function{QuantileUQ}{$D, \tau_i, \gamma$}
    \State $ b=1000$ \Comment{Bootstrap Number} 
    \State $n = \mathrm{nrow}(D)$ \Comment{Number of Samples}
    \For{$i$ = 1:$b$} \Comment{Bootstrap Loop}
        \Statex 
        \State $s_i = \text{{\myfont sample}}(n, \mathrm{replace=TRUE})$  \Comment{Bootstrap Indices}
        \State $D^* = D\left[s_i,\right]$ \Comment{Bootstrap Resample} 
        \State $R^* = \text{{\myfont scale}}(D^*, \mathrm{center = TRUE, scale = TRUE})$  \Comment{Standardize Resample} 
        \State $C^* = \text{{\myfont cor}}(R^*)$ \Comment{Bootstrap Correlation Matrix} 
        \State $J\left[i\right] =  \underset{p}{\arg\min} \biggl[ g(p; \tau_i, C^*) \biggl] $ \Comment{\eqref{EQ:Min}} \setstretch{2} 
    \EndFor{} \setstretch{1.25} 
    \State $\tau_{\gamma} = \text{{\myfont percentile}}(J, \gamma)$ \Comment{Percentile Bootstrap Confidence Interval}
    \State \textbf{Return}:  $\tau_{\gamma}$
    \EndFunction
    \end{algorithmic}
\end{algorithm}

\section{Tolerance Regions in Multivariate Statistics}\label{Tolerance_Regions}
\subsection{Traditional Tolerance Regions }\label{Elliptical_Tolerance}
\begin{sidewaysfigure}[h!] 
\includegraphics[width=\textwidth, keepaspectratio]{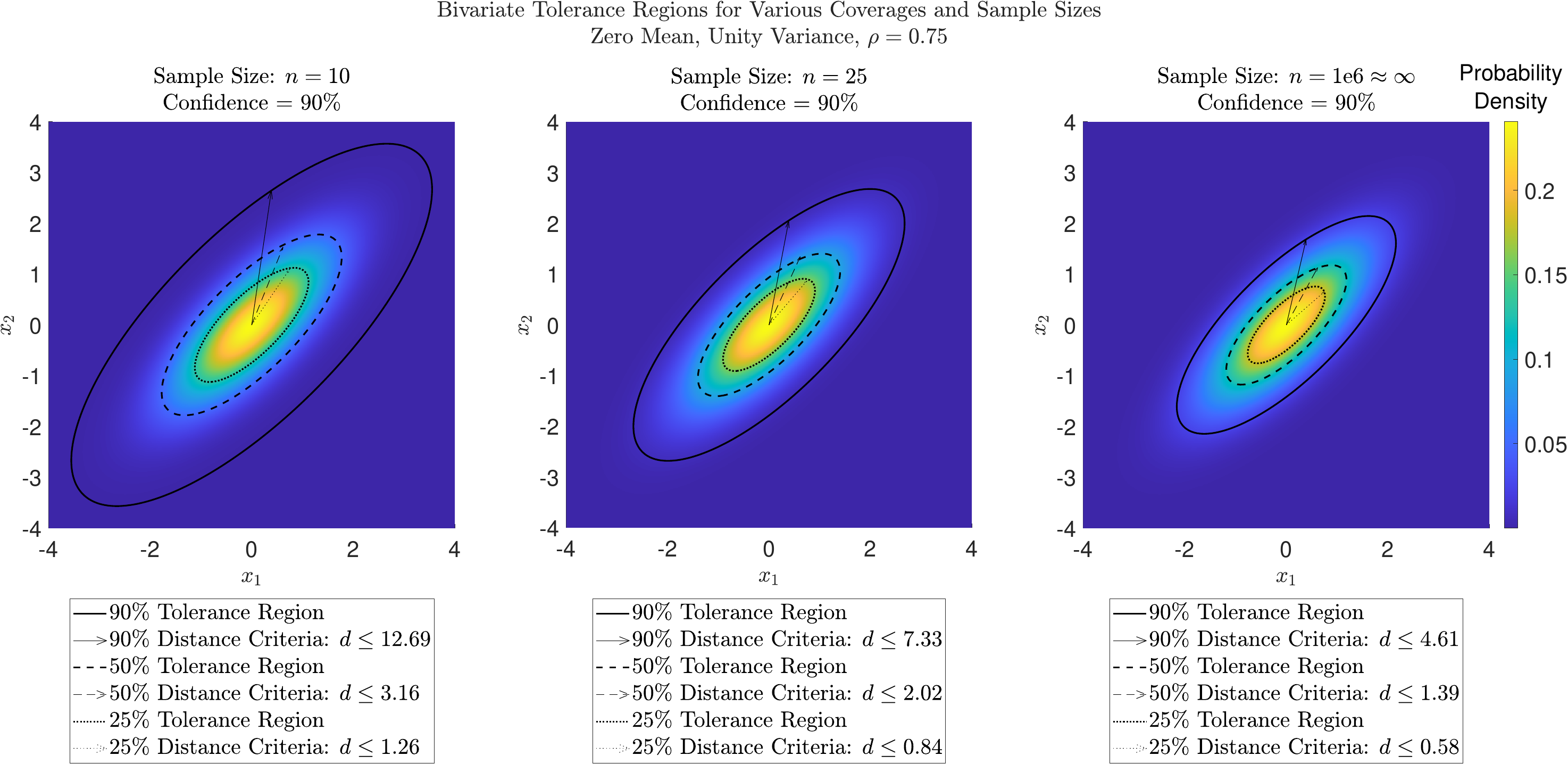}
\centering
\caption{Various examples of normal tolerance regions for a standard normal bivariate distribution with $\rho=0.75$. Three different proportion values are shown, $\beta = \left\{0.25, 0.50, 0.90 \right\} $ for three different sample sizes, $n = \left\{10, 25,  1\mathrm{e}6 \approx \infty \right\}$. All the tolerance regions plotted use a constant confidence value of 90\% and using algorithm 9.2 from \cite{krishnamoorthy_multivariate_2009}.}
\label{Plot:Tolerance_Regions}
\end{sidewaysfigure}

Traditional tolerance regions can be understood as a higher dimension generalization of the univariate two-sided tolerance intervals. Construction of traditional tolerance regions for a $q$-variate normal distribution is performed through Monte Carlo simulation from Algorithm 9.2 in \cite{krishnamoorthy_multivariate_2009} to estimate the threshold squared Mahalanobis distance, $r$. Figure \ref{Plot:Tolerance_Regions} depicts several examples of what traditional tolerance regions look like for a bivariate distribution with a zero vector mean, unity variance, and with a correlation of $\rho=0.75$. Three different proportion values, also known as coverage values, were chosen, $\beta \in \left\{0.25, 0.50, 0.90 \right\} $ for three different sample sizes, $n \in \left\{10, 25,  1\mathrm{e}6 \approx \infty \right\}$. The tolerance regions plotted use a constant confidence value of 90\%. Larger confidence values would increase the area of the elliptical tolerance region while smaller confidence values would decrease the area. As the sample size increases, the elliptical tolerance regions shrink due to increasing certainty of the multivariate distribution's parameters. As the sample size reaches infinity, or equivalently when the distribution's parameters are known, the confidence term has no importance as there is no uncertainty on the distribution's parameters. There is a lack of nomenclature as to what to call these elliptical tolerance regions when the distribution's parameters are known; going forward, the paper will refer to them as concentric elliptical quantiles. Elliptical tolerance regions imply the smallest region, centered at the mean, that encompasses at least a proportion $\beta$ of the population for a particular confidence. 

One can also interpret traditional tolerance intervals through the lenses of real analysis. Utilizing the bivariate distribution as an example, the joint probability density function $f(x_1, x_2)$ is:
\begin{align}
 f(x_1, x_2) = \frac{\exp{ \bigg( -\frac{1}{2}(\boldsymbol{x}-\boldsymbol{\mu})^T  \Sigma^{-1} (\boldsymbol{x}-\boldsymbol{\mu})}\bigg)}{\sqrt{(2 \pi)^2 \mathrm{det}(\Sigma)}},
\end{align}
where $ \mathrm{det}(\bullet)$ denotes the determinant operator and $\boldsymbol{\mu} = [\mu_1, \mu_2]$. One can write the tolerance region as,

\begin{gather}
\beta = \int \int_{\Omega}  f(x_1, x_2) \ dx_{1} \ dx_{2} :\label{EQ:TOL_UPPEReranceBeta}  \\ 
\Omega = \left\{ \boldsymbol{x}=(x_1, x_2) \  | \  h(\boldsymbol{x},\bar{\boldsymbol{x}}, S) \le r \right\},   \nonumber
\end{gather}
where $r$ is the threshold squared Mahalanobis distance. Traditional tolerance regions imply coverage integrated outwards from the mean, $\boldsymbol{\mu}$. This can be found directly in \eqref{EQ:Mahalanobis} of the form, $(\boldsymbol{x}- \bar{\boldsymbol{x}})$. Another interpretation of \eqref{EQ:TOL_UPPEReranceBeta} can be visualized as the third column of Figure \ref{Plot:Tolerance_Regions}, when $d \le r=4.61$, this implies $\beta=0.90$, therefore 90\% of the population resides within a squared Mahalanobis distance of less than or equal to $r=4.61$ units. Tolerance regions in $\mathbb{R}^2$ take the form of an ellipse, in $\mathbb{R}^3$ an ellipsoid, and for $\mathbb{R}^q : q > 3$, the tolerance region is a hyper-ellipsoid.

\afterpage{\clearpage}

\subsection{CDF Based Quantile Contours vs. Concentric Elliptical Quantiles for Bivariate Normal Distributions}\label{Comparison}

\begin{sidewaysfigure}[h] 
\includegraphics[width=\textwidth]{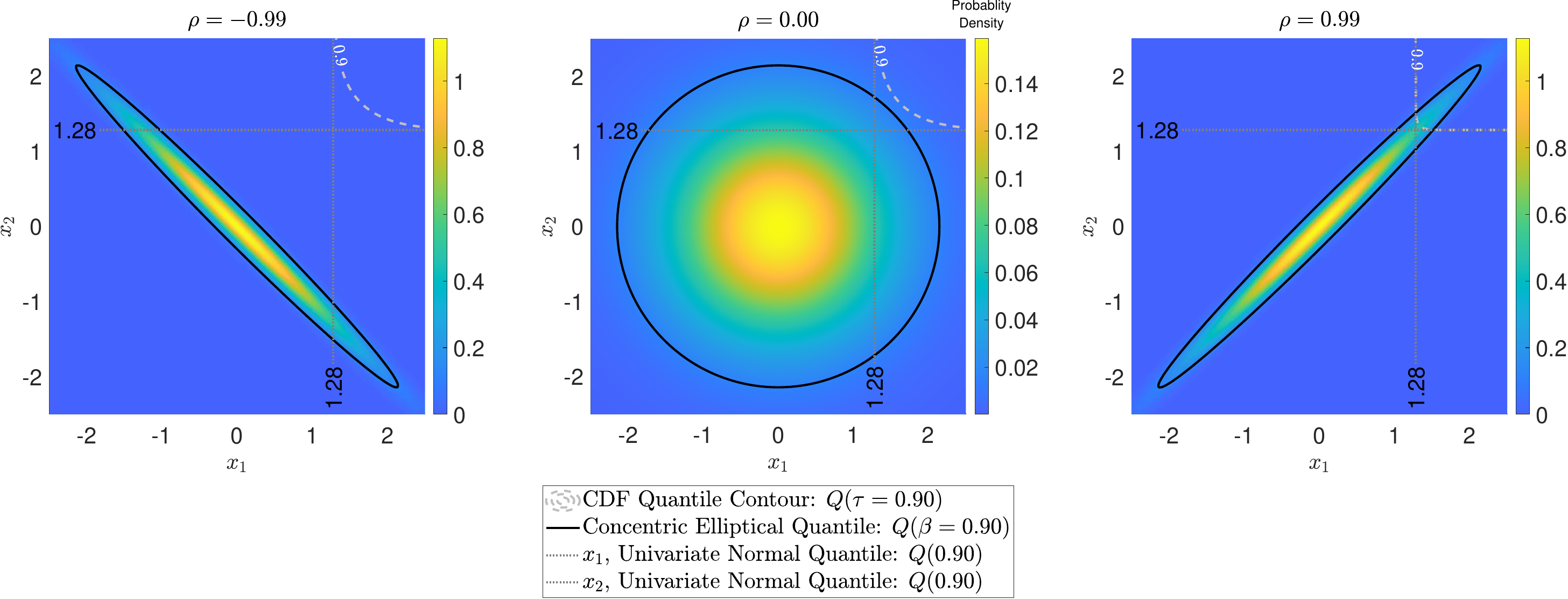}
\centering
\caption{Comparing concentric elliptical quantiles to CDF based quantile contours when the distribution parameters are known. A standard normal bivariate distribution with various correlation coefficients at, $\rho=\left\{-0.99, 0, 0.99 \right\},$ respectively are shown.}
\label{Plot:EllipticalQuantile_vs_QuantileContour}
\end{sidewaysfigure}

The easiest case for comparing concentric elliptical quantiles to CDF based quantiles occurs for the bivariate normal distribution.  Figure \ref{Plot:EllipticalQuantile_vs_QuantileContour} demonstrates how the two quantile methods compare for correlation coefficients at, $\rho \in \left\{-0.99, 0, 0.99  \right\}$, respectively for a standard normal bivariate distribution. When $\rho=-0.99$, the concentric elliptical quantile associated with $\beta=0.90$ are quite far from the CDF quantile contour associated with  $\tau=0.90$. An infinite number of points exists on an ellipse. Trying to select one or a few points on the ellipse when $\rho=-0.99$ for multivariate tolerance applications is quite challenging. Maximizing a value for one variate implies minimizing the value for another variate. For quality control applications, testing a component at many points on a multivariate quantile is cost prohibitive. In mechanical shock and vibration, testing a single component at more than one point on the quantile could lead to degradation of the component's mechanical performance by inadvertently fatiguing the component. When $\rho=0.99$, the concentric elliptical quantile associated with $\beta=0.90$ extends well beyond the CDF quantile contour associated with $\tau=0.90$ and even further past the CDF quantile contour associated with $\tau=0.95$ (not shown). For the case where the variates are uncorrelated, $\rho=0$, the ellipse is now a circle and is relatively close to the CDF quantile contours. It is quite evident by comparing the CDF quantile contours to the concentric elliptical quantiles that traditional tolerance regions do not consider the tails of the multivariate distributions. 

\afterpage{\clearpage}

\subsection{CDF Based Quantile Contours vs. Simultaneous Univariate Tolerance Intervals}\label{Comparison2}

When one doesn't know the distribution's parameters, Krishnamoorthy suggests to use a simultaneous univariate tolerance interval to estimate the extreme responses of a normal multivariate distribution \citep{krishnamoorthy_multivariate_2009}. A one-sided, upper, simultaneous univariate tolerance bound, is shown below:
\begin{align}
    \UT{T}_{\mathrm{simul}, \beta} \bigg(1- \frac{\alpha}{q}; \beta, n \bigg). \label{EQ:TOL_UPPER_SIMUL}
\end{align}
A simultaneous univariate tolerance bound is simply adjusting the confidence parameter, $\frac{\alpha}{q}$, by a Bonferroni correction by incorporating the number of variates, $q$, into a regular univariate tolerance interval. When the sample size approaches infinity, the simultaneous univariate tolerance bound converges to the standard univariate quantiles for each variate. While extremely efficient to compute, it doesn't provide a unique point on a multivariate quantile but rather a point such that the coverage of the multivariate distribution is at least $\beta$ for a particular confidence value.

Much of this paper up until now has been designed to build up a case for how correlation can significantly affect the CDF based quantiles for multivariate normal distributions. All the nuances for how correlation affect the quantiles, see Figure \ref{Plot:Bivariate_PDF_CDF} as an example, is lost by using the Bonferroni correction in the simultaneous univariate tolerance intervals. The simultaneous univariate tolerance intervals do provide a desired coverage but that particular point doesn't satisfy the definition of a point associated to the $\tau^{\mathrm{th}}$ quantile probability.  Therefore, there is a strong need for a new method that incorporates the correlation of the data, accounts for the tails of the multivariate distribution, while also maintaining the conservative properties that a point residing on the CDF based quantiles has. 

\section{A Need for a New Methodology}\label{CI_MV_Quantiles}
There are many instances where one is more concerned about extreme multivariate responses in contrast to what squared Mahalanobis distance encompasses a proportion of the population for a given confidence value. Therefore, one would want a CDF based quantile compared to an elliptical tolerance region. This is analogous to choosing a one-sided vs two-sided statistical interval in univariate statistics. 

Two main challenges exist when working with the quantile function \eqref{EQ:MV_Quantile} for multivariate normal distributions. The first is that an infinite number of vectors, $\boldsymbol{x}$, satisfy the condition  $\tau \le F(\boldsymbol{x})$. The second, even if there was a finite set of vectors that satisfied the condition above, no analytical solution is known for the quantile function \eqref{EQ:MV_Quantile} and therefore must be numerically estimated.

The following subsections describe two methodologies for calculating confidence intervals on multivariate normal quantiles given the challenges listed above. The first methodology is a computationally intense Monte Carlo meshing algorithm.  The goal of the first methodology is to estimate uncertainty on CDF based multivariate quantiles over a reasonable domain that captures an essential, finite, subset of vectors from the quantile as they extend towards their asymptotic limits. This methodology will be denoted simply as confidence intervals on multivariate quantiles. Since multivariate quantiles contain an infinite set of vectors, the goal of the second methodology is to estimate uncertainty on the vector within the set that has the highest probability density value.
\begin{align}
    \boldsymbol{v}(\tau) = \underset{ \boldsymbol{x}  \ : \  \boldsymbol{x} \  \in \   Q(\tau) }{\arg\max} \biggl[ f\bigl(Q(\tau)\bigl) \biggl] ,  \label{EQ:MaxPDF}
\end{align}
where $Q(\tau)$ is the set of vectors that are associated to the $\tau^\mathrm{th}$ quantile probability and $f(\boldsymbol{x})$ is the PDF of a $q$-variate normal distribution. The vector within the set associated to the CDF-based quantile, $Q(\tau)$, that has the highest probability density is given the novel term: \textit{critical point}. This methodology will be denoted simply as confidence intervals of critical points on multivariate quantiles. A critical point can be understood as the most ``likely" value on a CDF based quantile to occur. We present without proof that a unique critical point exists. When overlaying the CDF based quantiles with the PDF contours for multivariate distributions that are zero mean with unity variance, the CDF based quantiles are tangent to the PDF contours at the equicoordinate line for all possible correlations. Therefore, any translations from the equicoordinate line reduces the probability density.

\subsection{Standardizing Multivariate Data} \label{Standardizing}
The emphasis on standardizing multivariate data or multivariate distributions cannot be overstated for any of the methodologies presented herein. In univariate statistics, the process of transforming data such that it has a zero mean and unit variance is often called a z-transform, standardization, or auto-scaling. Commonly notated as:
\begin{align}
z_i = \frac{x_i - \bar{x}}{s},
\end{align}
where $z_i$ is the transformed observation, $x_i$ is the original observation, $\bar{x}$ and $s$ are the sample mean and sample standard deviation, respectively. Standardization for multivariate data takes an analogous matrix form and is denoted as  $\Psi(D)$,
\begin{align}
\mathcal{D} = \Psi(D) := (D- \bar{\boldsymbol{x}}) \bigg( {\sqrt{\mathrm{diag}\big({\mathrm{diag}({S})} \big)}} \bigg)^{-1} \ ,
\end{align}
where $\mathcal{D}$ is the transformed multivariate data and $D \in \mathbb{R}^{ (n \times q)}$ is a matrix of data originating from a $q$-dimensional multivariate distribution with $n$ samples. The diagonal matrix where the elements along the diagonal are the sample variances is denotes as $ {\mathrm{diag}\big({\mathrm{diag}({S})} \big)}$ and the sample covariance matrix is denoted as $S$. The affine transformation back to the original domain is thus, $D = \Psi^{-1}(\mathcal{D})$. Standardization in multivariate statistics preserves the correlation between the multivariate data while transforming the data such that is has a zero-mean vector. The covariance matrix of the transformed data now has unity variances along diagonal of it, thus making the covariance matrix also the correlation matrix. 

Standardization is utilized in constructing confidence intervals on multivariate normal quantiles. When standardizing the multivariate data, only one numerical study is required to determine the optimal mesh step size to avoid graphical aliasing of the contours or iso-surfaces, see \Cref{Method1}. Constructing confidence intervals for critical points on multivariate quantiles also requires the multivariate data to be standardized for the numerical method to work, see \Cref{Method2}. 

\subsection{Method 1 - Confidence Intervals on Multivariate Normal Quantiles} \label{Method1}

The purpose of this section to describe a methodology for constructing confidence intervals on CDF based multivariate normal quantiles. The need for this methodology occurs when the multivariate distribution's parameters are \textit{not} known and when one requires more than just a single critical point that resides on the quantile. Several considerations must be made to properly mesh the support of the multivariate distribution. One, what are the bounds of the mesh that is being tessellated. Two, what granularity is required to properly discretize the support. To address the first point, since the multivariate data is standardized, as is discussed in \Cref{Standardizing}, we suggest that the bounds over the domain are, $\Omega \subset \mathbb{R}^q = \left\{ -4 \le x_i \le 4  \right\} : i = {1, \ldots, q}$ due to the low probability of values outside of this domain. A Bonferroni lower limit for the minimum probability over the  $q$-dimensional space is calculated below for  univariate distributions up to trivariate distributions, $q=\left\{ 1, 2, 3 \right\}$,
 \begin{align}
1 - \bigg[q\bigg(1 - F(4 \times \boldsymbol{1}_q) + F(-4 \times \boldsymbol{1}_q)\bigg) \bigg] = \left\{0.999936, 0.999873, 0.999810\right\}, \label{EQ:tessellatedProb}  
\end{align}
where $\boldsymbol{1}_q$ denotes a one vector of size $q$ and $ F(\bullet)$ is still the CDF function for a $q$-variate normal distribution. From \eqref{EQ:tessellatedProb}, one can see that the domain being tessellated covers most of the codomain of $F$.

\begin{figure}[H]
\centering
\subfigure[CDF quantile contour from a bivariate distribution with a correlation of, $\rho=-0.99$.]{%
\resizebox*{7cm}{!}{\includegraphics{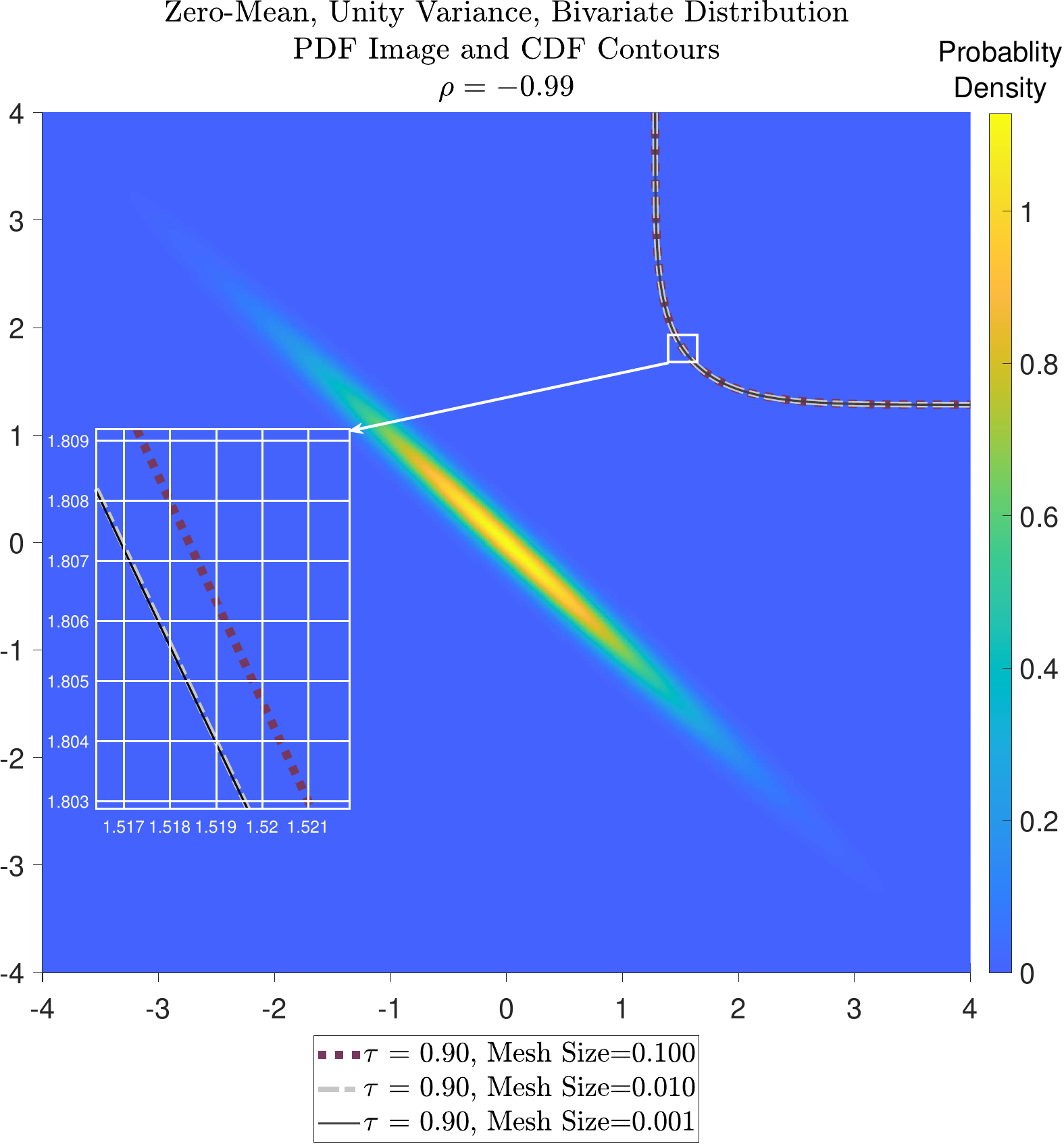}} \label{Plot:CDF_Mesh_A}}
\subfigure[CDF quantile contour from a bivariate distribution with a correlation of, $\rho=0.99$.]{%
\resizebox*{7cm}{!}{\includegraphics{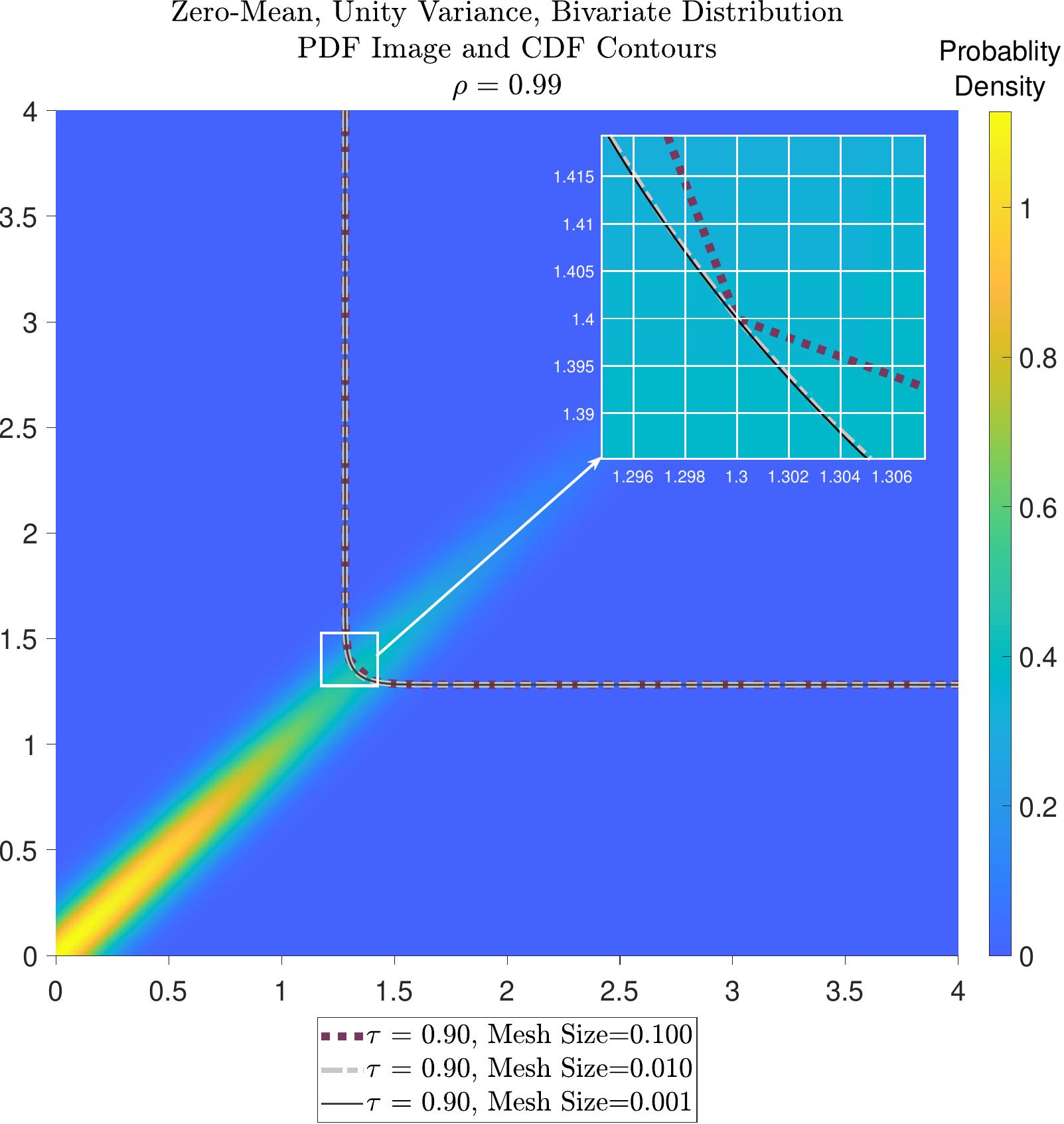}}\label{Plot:CDF_Mesh_B}}\hspace{5pt}
\caption{CDF quantile contours represent mesh step sizes with values of, $\left\{0.001, 0.01, 0.1   \right\}$, for solid, dashed, and dotted,  curves respectively. Both plots are for a bivariate quantile probability value of $\tau = 0.90$. The solid and dashed lines are superimposed onto each other representing that the mesh size is sufficiently small for the bivariate case.} \label{Plot:Bivariate_CDF_Mesh}
\end{figure}

To address the second question regarding how finely one discretizes the support, a study was performed using three different mesh step sizes with values of, $\left\{0.001, 0.01, 0.1 \right\}$, shown in Figure \ref{Plot:Bivariate_CDF_Mesh}. The study shows little to no aliasing of the CDF based quantile contours for mesh sizes less than 0.01 units for precision at three decimal places. Even with a mesh size of 0.1 and a high correlation coefficient of $\rho=0.99$, the CDF based quantile contour is nearly indistinguishable to the contours meshed at a smaller size. When graphical aliasing does occur, it is at areas of strong curvature, high positive correlation values, and  where the PDF is highest. The authors recommend using a mesh of 0.01 for constructing confidence intervals for bivariate normal distributions and a mesh size of 0.1 for trivariate normal distributions. The computational burden of meshing the support for a trivariate normal distribution is offset using a larger mesh size of 0.1 but at the expense of accuracy and graphical aliasing.   

Algorithm \ref{ALGO:CI_MVQuantile} provides the methodology to calculate  confidence intervals on trivariate normal quantiles. The algorithm can generalize to higher dimensions, tessellation is extended over the additional dimensions. Nevertheless, the computation burden of tessellating in higher dimensions grows exponentially: $m^q$. Similarly, the algorithm can generalize to the bivariate case and interpolation is also modified from 3 dimensions to $q=2$ dimensions. The algorithm provides two different options for bootstrapping: a parametric and nonparametric method. Parametric bootstrapping utilizes the original sample mean and sample covariance to randomly generate multivariate data of the same size as the original. \cite{efron_jackknife_2016} has suggested that a parametric family, like a multivariate normal distribution, can act as regularizers, smoothing out the raw data and de-emphasizing outliers. A nonparametric bootstrap method is also provided which imposes fewer constraints and assumptions on the underlying data. When working with the nonparametric bootstrap approach, one must be cognizant that resampling data with low initial sample sizes, $n$, can lead to degenerate matrices. Thus, one must either resample again until the problem is alleviated or simply use the parametric approach which avoids these issues.    

The argument, \textit{interpolate}, found in Algorithm \ref{ALGO:CI_MVQuantile}, is a Boolean data type. When the variable is true, cubic interpolation occurs over the $q$-dimensional space, for the percentiles of the tessellated CDF mesh by an upsampling factor of $\xi=2$. Interpolation on the tessellated mesh prior to final contour method provides additional accuracy on the final contours that would be originally inhibited by the memory required to store the CDF tensor at such fine granularity. Interpolation has precedence in previous multivariate statistical CDF work by \cite{langrene_fast_2021} who utilized it in their fast multivariate empirical CDF calculations.

\begin{algorithm}[H]
\caption{ - MATLAB pseudo code for a function that calculates confidence intervals for trivariate normal CDF based quantiles. $D \in \mathbb{R}^{ (n \times 3)}$ is a matrix of data originating from a $3$-dimensional multivariate distribution with $n$ samples. $\tau$ is the desired multivariate quantile probability value. $\gamma$ are the $\gamma^\mathrm{th}$ percentile values used to construct the simple percentile bootstrap confidence intervals on the tessellated CDF mesh denoted as, $Z_{\gamma}$. The argument, \textit{style}, controls whether to perform parametric and nonparametric bootstrapping. The argument, \textit{interpolate}, is a Boolean data type that when true, interpolates the percentiles of the tessellated CDF mesh by an upsampling factor of $\xi$. To denote that interpolation was performed on the tessellated CDF mesh, the breve symbol is used, $\breve{Z}_{\gamma}$. The hat symbol on $\hat{Q}_{\gamma} $ denotes that this is an estimated subset of $Q(\tau)$, since $Q(\tau)$ is an infinite set of vectors. Lastly, the under-bar accent in, $\underaccent{\bar}{\hat{Q}}_{\gamma}$, denotes that the operation was performed in the standardized domain. The inverse, affine, transformation of the quantile back to the original domain of the data is simply denoted without the under-bar.}
\label{ALGO:CI_MVQuantile}
    \begin{algorithmic}[1]
    \Require $0 < \tau < 1, \quad 0 < \gamma < 1, \quad$
    \Function{TrivariateQuantile}{$D, \tau,\gamma$, \textit{style}, \textit{interpolate}}
    \State $ b=1000$ \Comment{Bootstrap Number}  
    \State meshStep = 0.1
    \State $[n, q] =\text{{\myfont size}}(D)$ \Comment{Number of Samples and Variates}
    \State $\mathcal{D} = \text{{\myfont normalize}}(D, \mathrm{``zscore"})$  \Comment{Standardize Data}  
    \State $S = \text{{\myfont cov}}(\mathcal{D})$  \Comment{Covariance/Correlation Matrix} 
    \State x = [-4 : meshStep : 4]
    \State y = [-4 : meshStep : 4]
    \State z = [-4 : meshStep : 4]
    \State [X,Y,Z] =  \text{{\myfont meshgrid}}(x,y,z)
    \State XYZ = [X(:), Y(:), Z(:)]
    \State MC = \text{{\myfont nan}}(length(y),length(x),length(z),b) \Comment{Pre-Allocate nan tensor}
    \For{$i$ = 1:$b$} \Comment{Monte Carlo Loop}
        \If{\textit{style} == `parametric'} \Comment{Parametric Bootstrap}
            \State $\mathcal{D}^* =\text{{\myfont mvnrnd}}(n, \mathrm{mu}=\boldsymbol{0}, \mathrm{Sigma}=S)$   \Comment{Parametric Resample}
        \ElsIf{\textit{style} == `nonparametric'} \Comment{Nonparametric Bootstrap}
            \State $\mathcal{D}^*= \text{{\myfont datasample}}(\mathcal{D}, n, \mathrm{replace=TRUE})$  \Comment{Nonparametric Resample} 
        \EndIf
        \State $\bar{\boldsymbol{x}}^* = \text{{\myfont mean}}(\mathcal{D}^*, \mathrm{DIM}=1)$  \Comment{Bootstrap sample Mean} 
        \State $S^* =  \text{{\myfont cov}}(\mathcal{D}^*)$  \Comment{Bootstrap sample covariance}   
        \State $Z^* = \text{{\myfont mvncdf}}(\mathrm{XYZ}, \bar{\boldsymbol{x}}^*, S^*)$  \Comment{Bootstrap CDF}     
        \State $\mathrm{MC}(:, :, :, i) = Z^*$;             \Comment{Store Bootstrap CDF}
    \EndFor{}
    \State $Z_{\gamma} = \text{{\myfont percentile}}(\mathrm{MC}, \gamma,  \mathrm{DIM}=q+1)\textbf{} $ \Comment{percentile the CDF Mesh} 
    \If{\textit{interpolate} == TRUE}
    \State $ [\breve{Z}_{\gamma}, \mathrm{x}, \mathrm{y}, \mathrm{z}] =  \text{{\myfont interp3}}(Z_{\gamma},  \mathrm{x},  \mathrm{y}, \mathrm{z}, \mathrm{\xi}=2)$
    \State $\underaccent{\bar}{\hat{Q}}_{\gamma} = \text{{\myfont contourc}}(\mathrm{x}, \mathrm{y}, \mathrm{z}, \breve{Z}_{\gamma}, \tau)$ \Comment{Extract Contours} 
    \Else
    \State $\underaccent{\bar}{\hat{Q}}_{\gamma} = \text{{\myfont contourc}}(\mathrm{x}, \mathrm{y}, \mathrm{z}, Z_{\gamma}, \tau)$ \Comment{Extract Contours} 
    \EndIf
    \State $ {\hat{Q}}_{\gamma} = \Psi^{-1}(\underaccent{\bar}{\hat{Q}}_{\gamma})$ \setstretch{1.25}  \Comment{Inverse Transform to Original Domain} 
    \State \textbf{Return}: ${\hat{Q}}_{\gamma}$ \setstretch{1.25} \Comment{Percentile Bootstrap Confidence Interval}
    \EndFunction
    \end{algorithmic}
\end{algorithm}


\begin{figure}[H]
\centering     

\subfigure[Pictograph describing how to generate bivariate normal CDF based quantile contours. Each bootstrap within the Monte Carlo simulation will create a tessellated grid with dimensions $m \times m$ for which the CDF is calculated over. From Algorithm \ref{ALGO:CI_MVQuantile}, the support is tessellated mesh -4 to 4 with a mesh step size of 0.01, thus $m=801$. After all $b$ Monte Carlo simulations have been performed, the CDF grids are now a tensor with dimensions $m \times m \times b$. The $\gamma^{\mathrm{th}}$ percentile is performed along the third dimension of the tensor, resulting  back to a final dimension $m \times m$. The $\gamma^{th}$ percentile provides uncertainty on the CDF values from all the various multivariate bootstrapped data. It is important to note that the $\gamma^{\mathrm{th}}$ percentile is over the tessellated CDF tensor, which is not to be confused with the $\tau^{\mathrm{th}}$ quantile probability which refers to the quantile probability associated with the multivariate normal distribution. ]{\label{Plot:MC_Quantiles_A}\includegraphics[page=1, width= 0.95 \linewidth]{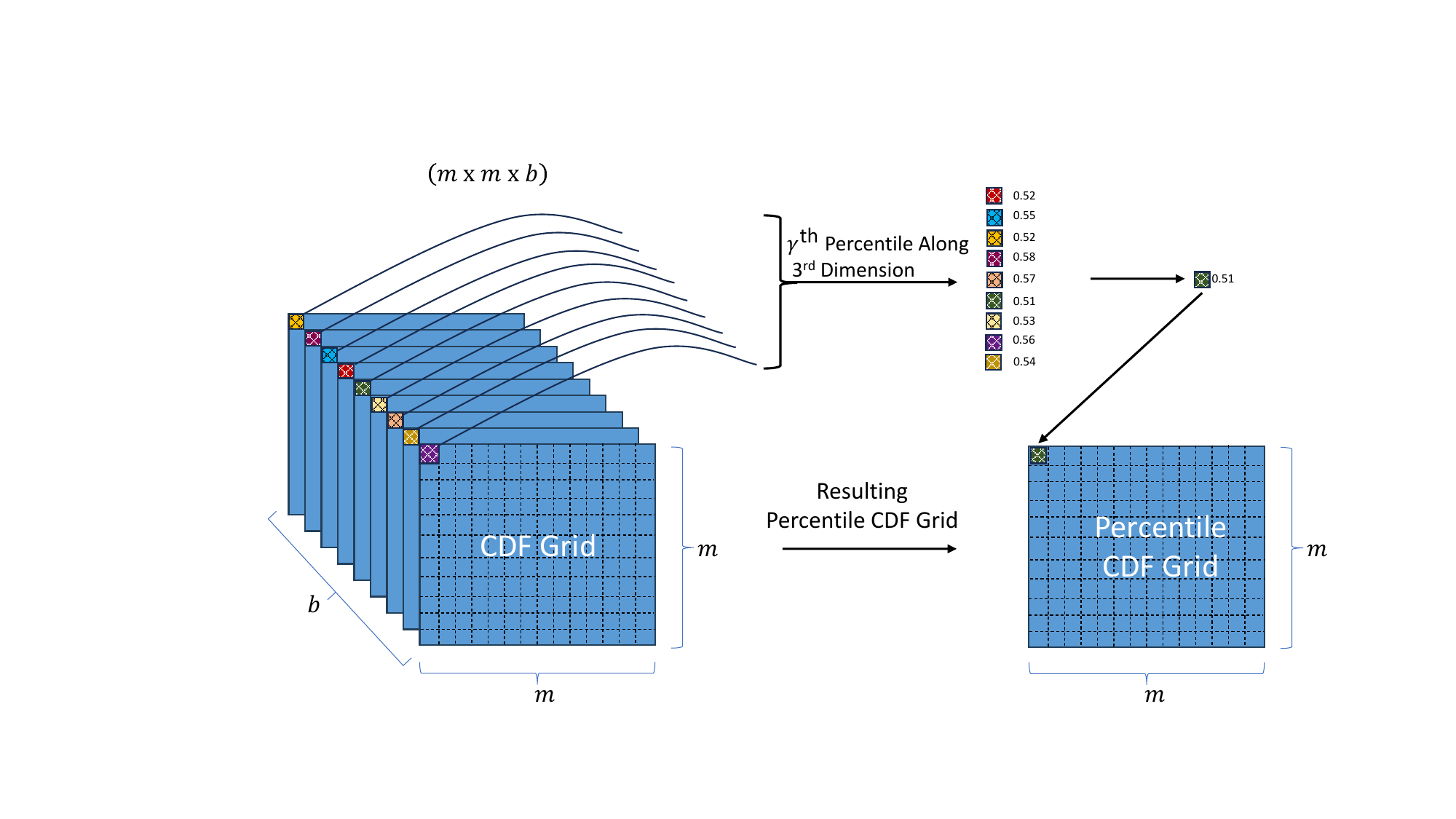}}

\subfigure[Once the $\gamma^{\mathrm{th}}$ percentile is computed over the tessellated CDF tensor from \ref{Plot:MC_Quantiles_A}, the contours associated to the $\tau^{\mathrm{th}}$ quantile probability can be estimated. The pictograph above provides a rudimentary example for a multivariate normal quantiles associated with $\tau=0.90$ highlighted in black with white colored font. The uncertainty associated with the set of vectors comprising a multivariate CDF quantile is denoted as ${\hat{Q}}_{\gamma}(\tau)$. The blank spots between the bottom left and top right of the grid would represent monotonically increasing CDF values. A similar procedure can be performed for trivariate normal distributions, however, the quantile associated for ${\hat{Q}}_{\gamma}(\tau)$ is an iso-surface from a tessellated cube that with dimensions $m \times m  \times m$.]{\label{Plot:MC_Quantiles_B}\includegraphics[page=1, width= 0.95 \linewidth]{How_to_Generate_a_Quantile_of_a_CDF.pdf}}

\caption{Pictographs describing the process of generating confidence intervals for CDF based bivariate normal quantiles, ${\hat{Q}}_{\gamma}(\tau)$. The methodology is similar for higher dimensions such as trivariate normal quantiles, but with much higher computational burden for an increase in dimensionality.}
\end{figure}

\begin{figure}[H] 
\includegraphics[width= \textwidth]{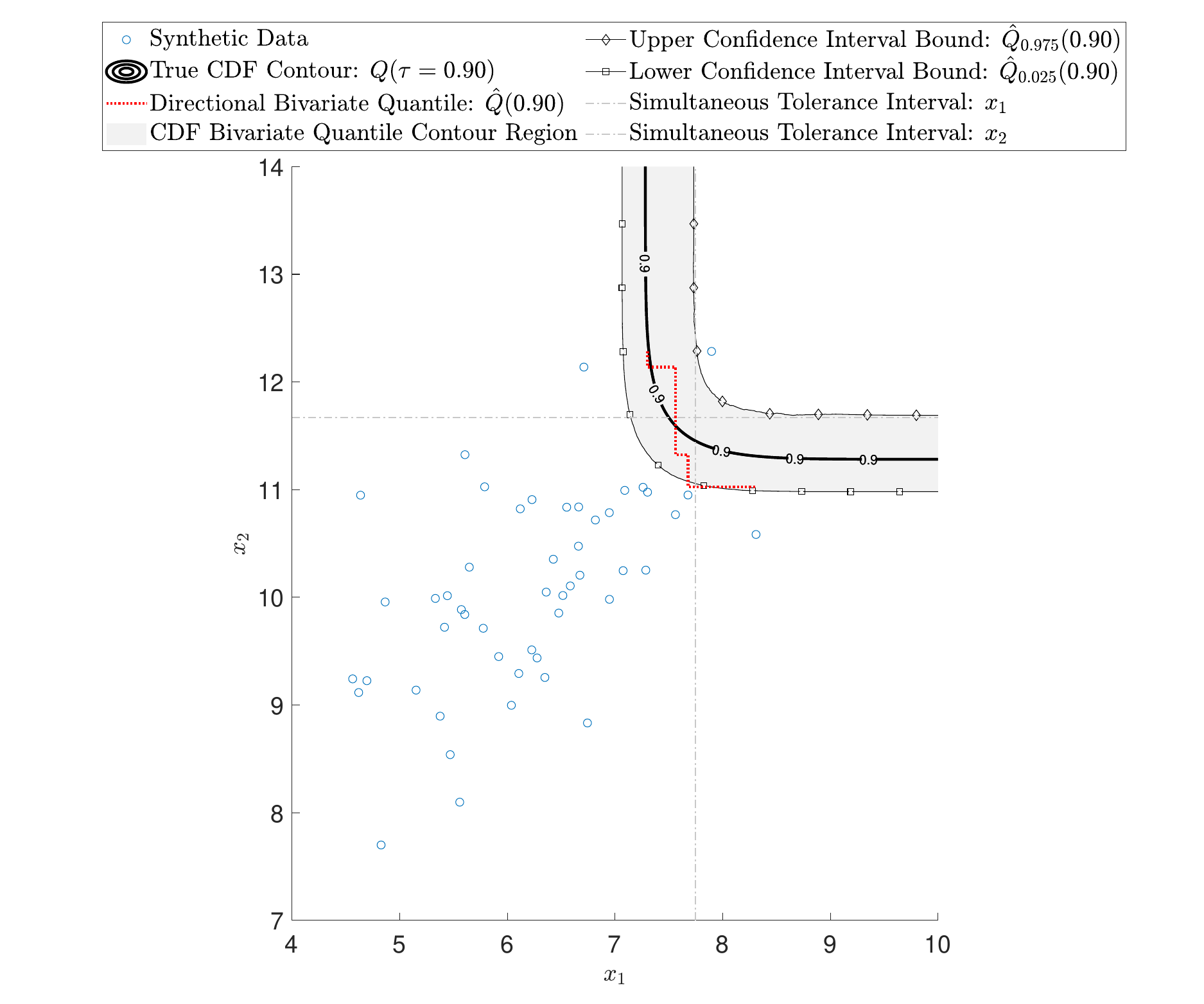}
\centering
\caption{Bivariate normal distribution example comparing the directional bivariate quantile methodology and the present methodology described above for constructing confidence intervals on multivariate normal quantiles. The true contour, $Q(\tau=0.90)$, is shown as a thick black line. The directional bivariate quantile from \cite{klein_directional_2020} is shown as a red dotted line. The methodology described above for constructing confidence intervals on multivariate normal quantiles, ${\hat{Q}}_{\gamma}(\tau)$, are shown as the shaded grey region for the quantile probability of $\tau=0.90$ using a two-sided confidence interval of 95\%. To generate the quantiles, $b=5000$, was used as the number of Monte Carlo simulations followed by using the BCa bootstrap confidence interval method. The upper confidence interval, ${\hat{Q}}_{0.975}(\tau)$, is shown as a line with diamond markers, while the lower confidence interval, ${\hat{Q}}_{0.025}(\tau)$, is shown as a line with square markers.}
\label{Plot:DirectionalBivariate_VS_CDFContours}
\end{figure}

In Figure \ref{Plot:DirectionalBivariate_VS_CDFContours} we compare the bivariate version of algorithm \ref{ALGO:CI_MVQuantile} to the only current methodology for generating multivariate quantiles, the latter is only applicable in the bivariate case \citep{klein_directional_2020}. Fifty random bivariate vectors were sampled from the same bivariate distribution to create a synthetic data set, $D \in \mathbb{R}^{(50 \times 2)}$, shown as blue circles. The $i^{\mathrm{th}}$ random vector are sampled from a bivariate normal distribution with parameters, $D_i \sim \mathcal{N} \bigg( \boldsymbol{\mu}=[6, 10]), \Sigma=\begin{bmatrix} 1 & 0.50 \\ 0.50 & 1 \end{bmatrix} \bigg)$. The true contour, $\tau=0.90$, is shown as a thick black line. The directional bivariate quantile from \cite{klein_directional_2020} is shown as a red dotted line. The methodology described above for constructing confidence intervals on multivariate normal quantiles is shown as the shaded grey region for the quantile probability of $\tau=0.90$ using a two-sided confidence interval value of 95\%. To generate the quantiles, $b=5000$, was used as the number of Monte Carlo simulations followed by using the bias corrected and accelerated method (BCa) bootstrap confidence interval method. The BCa bootstrap confidence interval method was utilized instead of the basic percentile method because it generally has smaller coverage error \citep{davison_bootstrap_1997, carpenter_bootstrap_2000}. The algorithms presented here use the basic percentile method because implementing the (BCa) methods are more complicated. The upper confidence interval is shown as a line with diamond markers, while the lower confidence interval is shown as a line with square markers. Figure \ref{Plot:DirectionalBivariate_VS_CDFContours} demonstrates that even with 50 samples, the current directional bivariate method provides a rough empirical approximation to the true bivariate contour but with no measure of uncertainty. The methodology described here provides not only uncertainty on the distribution's quantile but also the smoothness one expects. All the quantiles shown in Figure \ref{Plot:DirectionalBivariate_VS_CDFContours} asymptotically extend towards infinity with respect to each variate, yet one can see that domain for which the support is meshed captures the essence of the underlying bivariate quantiles. 

 The computational time on a 12th Gen Intel® Core™ i9-12950HX 2.30 GHz processor is about 189 seconds when using 5000 bootstrap resamples for the present methodology compared to about 0.90 seconds for the directional bivariate quantile method. The smoothness of the calculated quantile, the extended domain for which the quantile covers, and the measure of uncertainty on the quantile does come at a computational cost of approximately two orders of magnitude, nonetheless, the overall computational time is still reasonable for most applications. Computation time for iso-surfaces, representing quantiles for trivariate distributions, require significantly more computation time of approximately 6 hours for 1000 bootstrap resamples. As of to-date, algorithm \ref{ALGO:CI_MVQuantile} is the only method for calculating confidence intervals for quantiles associated with trivariate normal distributions. 
 
\subsection{Critical Point Selection} \label{Critical_Point}
 
 The single point on a multivariate quantile that has the maximum relative likelihood, or equivalently, highest probability density of occurring, is known as the critical point. The critical point will be denoted as the bold vector, $\boldsymbol{v}(\tau)$. These points can be estimated from the finite subset of vectors comprising the multivariate quantiles using a variety of techniques. If the data has been standardized, the critical point can be represented as $\boldsymbol{v} = \boldsymbol{1}_{q} v$ where $v$ is the equicoordinate value of the CDF based quantile, i.e., $v=x_1=\ldots=x_q$. In $\mathbb{R}^2$ the intersection of the equation $y=x$ with the CDF based quantile, represented as a contour, is the critical point. In $\mathbb{R}^3$ the intersection of the parametric equation:
 \begin{align}
    \begin{cases}
        x(t)=t\\
        y(t)=t\\
        z(t)=t
    \end{cases},
 \end{align}
with the CDF based quantile, represented now as an iso-surface, is the critical point.   

\begin{figure}[H] 
\includegraphics[width=\textwidth]{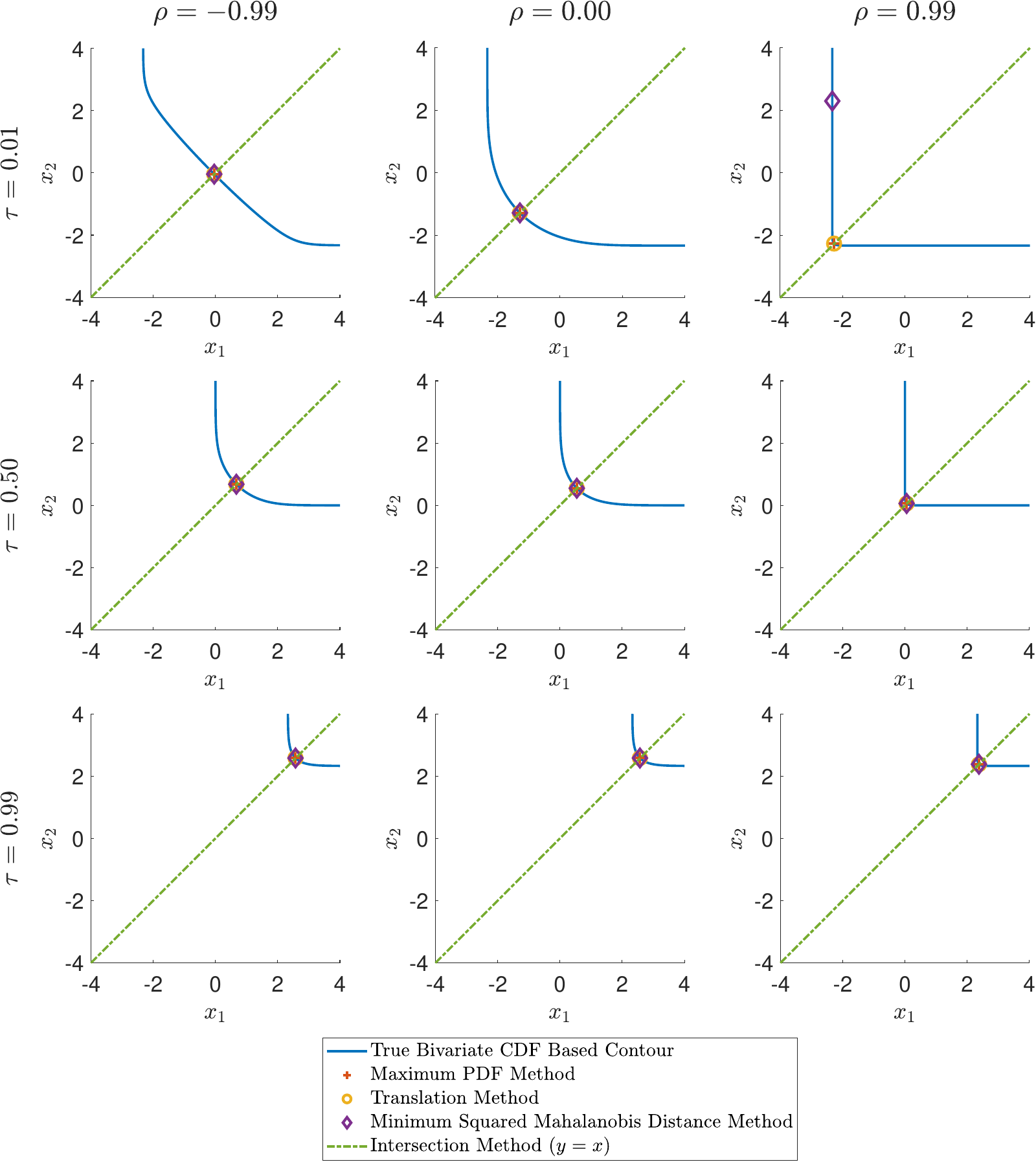}
\centering
\caption{Critical points, $\boldsymbol{v}(\tau)$, from a standardized bivariate normal distribution for three different quantile probabilities, $\tau = \left\{ 0.01, 0.5, 0.99 \right\}$, shown along the $x_2$ axis, and three different correlation values, $\rho = \left\{-0.99, 0, 0.99 \right\}$, shown along the $x_1$ axis.}
\label{Plot:BivariateCriticalPointMethods}
\end{figure}

Figure \ref{Plot:BivariateCriticalPointMethods} demonstrates the critical points, $\boldsymbol{v}(\tau)$, from a standardized bivariate normal distribution for three different quantile probabilities, $\tau = \left\{ 0.01, 0.5, 0.99 \right\}$, and three different correlation values, $\rho = \left\{-0.99, 0, 0.99 \right\}$. Since the distribution parameters are known for this example, one can select the vector that is associated with the largest probability density which is simply the vector from the quantile that has the maximum PDF value. Similarly, one can translate the desired CDF quantile from data that has been standardized into the first quadrant, followed by finding the vector that minimizes the $L^2$ norm. Translating the CDF quantiles into the first quadrant, or in higher dimensions orthant, forces all of the elements within the set of vectors that comprises a quantile to be positive. One can also see that the vector on the multivariate quantile that minimizes the squared Mahalanobis distance is generally the critical point but not always as for the case when $\tau=0.01$ and $\rho=0.99$, shown as a the purple diamond marker in the top right plot in Figure \ref{Plot:BivariateCriticalPointMethods}. Regardless of the method used to calculate the critical point, if the data is standardized, any of the methods, besides the minimum squared Mahalanobis distance, yield the same answer. When the parameters of the distribution are not known, either the translation or intersection method can be used when working with a particular quantile. To denote uncertainty on the critical points, $\boldsymbol{v}_\gamma(\tau)$, denotes the $\gamma^{\mathrm{th}}$ percentile on a critical point from any of the bootstrap confidence intervals associated for the $\tau^{\mathrm{th}}$ multivariate quantile probability.

\subsection{Method 2 - Confidence Intervals of Critical Points on Multivariate Quantiles} \label{Method2}
Under certain circumstances, one may not need a finite, subset of vectors for the multivariate quantiles but rather just the critical point for each quantile. When this is the case, one can substantially speed up the computation time by directly calculating only the critical point for a particular quantile, which is the topic for this section. One can utilize the {\myfont mvtnorm} library in the R language which provides a function called {\myfont qmvnorm} to help compute the critical values for multivariate normal distribution.

When the parameters of the multivariate normal distribution are known, thus confidence is irrelevant, one must first calculate the correlation matrix, $C$, from the known covariance matrix, $C = R  \Sigma R : R=\bigg(\sqrt{\text{diag}(S)}\bigg)^{-1}$. One then utilizes {\myfont qmvnorm} with the known correlation matrix to solve for the equicoordinate value, $v$, for a standard multivariate normal distribution. When the multivariate normal distribution is standardized, zero mean and unity variance, the equicoordinate vector can then be transformed to any arbitrary domain for all multivariate normal distributions with known distribution parameters, $ {\boldsymbol{v}} = \Psi(v) = \bigg(\boldsymbol{1}_{q} v \times  \sqrt{\text{diag}(\Sigma)} \bigg) + \boldsymbol{\mu}$ . 

When the parameters of the multivariate normal distribution are \textit{not} known, a Monte Carlo bootstrap procedure can be performed on the data. Algorithm \ref{ALGO:CI_CriticalPoints} provides an example of how to construct confidence intervals on the critical points on multivariate quantiles. Regardless of whether the parameters of the multivariate normal distribution are known or not, the calculations for the equicoordinate vector should be performed on a standard multivariate normal distribution. The critical point is then constructed by transforming the equicoordinate value to the domain of any arbitrary multivariate normal distribution.

\begin{algorithm}[H]
\caption{ - R pseudo code for calculating confidence intervals on the critical points on multivariate quantiles, $\hat{\boldsymbol{v}}_{\gamma}(\tau)$. $\gamma$ are the $\gamma^\mathrm{th}$ percentile values used to construct the simple percentile bootstrap confidence intervals on the critical points for a particular quantile probability, $\tau$. $D \in \mathbb{R}^{ (n \times q)}$ is a matrix of data originating from a $q$-dimensional multivariate distribution with $n$ samples.}
\label{ALGO:CI_CriticalPoints}
    \begin{algorithmic}[1]    
    \Require $0 < \tau < 1, \quad  0 < \gamma < 1$
    \Function{CriticalPointUQ}{$D, \tau, \gamma$}
    \State $ b=1000$ \Comment{Bootstrap Number} 
    \State $n = \mathrm{nrow}(D)$ \Comment{Number of Samples}
    \For{$i$ = 1:$b$} \Comment{Bootstrap Loop}
        \Statex 
        \State $s_i = \text{{\myfont sample}}(n, \mathrm{replace=TRUE})$  \Comment{Bootstrap Indices}
        \State $D^* = D\left[s_i,\right]$ \Comment{Bootstrap Resample} 
        \State $\mathcal{D}^* = \text{{\myfont scale}}(D^*, \mathrm{center = TRUE, scale = TRUE})$  \Comment{Standardize Resample} 
        \State $C^* = \text{{\myfont cor}}(\mathcal{D}^*)$ \Comment{Bootstrap Correlation Matrix} 
        \Statex
        \State $v^* = \text{{\myfont qmvnorm}}(\tau,  \mathrm{corr =}C^*,  \mathrm{tail = ``lower.tail"})$ 
        \Statex
        \State $\mu^* = \text{{\myfont apply}}(D^*, 2, \text{{\myfont mean}})$ \Comment{Bootstrap mean}
        \State $\Sigma^* = \text{{\myfont apply}}(D^*, 2, \text{{\myfont sd}})$ \Comment{Bootstrap covariance}
        \State $q = \text{{\myfont dim}}(D^*)[2]$ \Comment{Number of Variates}
        \State $V^*[i,] = \big(\text{{\myfont rep}}(v^*,q ) \times \Sigma^*\big) + \mu^* $ \Comment{Critical Point in Original Domain}
    \EndFor{} \setstretch{1.25} 
    \State $\hat{\boldsymbol{v}}_{\gamma}(\tau) = \text{{\myfont percentile}}(V^*, \gamma)$ \Comment{Percentile Bootstrap Confidence Interval}
    \State \textbf{Return}:  $\hat{\boldsymbol{v}}_{\gamma}(\tau)$
    \EndFunction
    \end{algorithmic}
\end{algorithm}

\section{Results}\label{Results}
\subsection{Numerical Simulation}\label{Numerical_Simulation}
\subsubsection{Algorithm 1 - Characterizing Uncertainty on the Multivariate Quantile Probability}\label{Numerical_Simulation_Algo1}
A numerical simulation was performed to show the efficacy of Algorithm \ref{ALGO:UQ_JointQuantile}. Recall that Algorithm \ref{ALGO:UQ_JointQuantile} characterizes the uncertainty on the multivariate quantile probability, $\tau$, by comparing to concurrent univariate quantile probabilities, $\tau_{1} = \ldots =\tau_{q}$. Two variables were investigated, the number of variates, $q$, and, $n$, the sample size of the data. The number of variates ranged from 2 to 4 (inclusively), $2 \le q \le 4$, and three different sample sizes were utilized in the simulation, $n \in \left\{  30, 300, 3000 \right\}$, thus leading to 9 combinations to test. A hundred Monte Carlo simulations were performed for each combination of variables listed in Table \ref{TBL:QuantileUQ}. For each Monte Carlo simulation, a random correlation matrix, $C_q$,was generated using the $C$-vine method from \cite{lewandowski_generating_2009}. The $C$-vine method requires sampling from a beta distribution which controls how much the distribution of partial correlations is concentrated around ±1 using just one parameter, $c$. For the context of the simulation, $c=2$, which provided correlation matrices with fairly large off-diagonal values. A random mean vector of dimension $q$ was sampled from a continuous uniform distribution: $\boldsymbol{\mu} \sim \boldsymbol{1}_q  \ \mathcal{U}(0, 100)$, where $\mathcal{U}(a, b)$ denotes a continuous uniform distribution over the support $\left[a, b \right]$. For each Monte Carlo simulation, data was randomly sampled from a multivariate normal distribution, $D \sim \mathrm{mvnrnd}(\boldsymbol{\mu}, C_q, n)$, where $D$ denotes a matrix of $n$ random vectors chosen from the same multivariate normal distribution, with mean vector $\boldsymbol{\mu}$ and correlation matrix $C_q$. 

Within each of the one hundred Monte Carlo simulations, the true joint coverage, $ \tau_{\mathrm{true}}$, was calculated using \eqref{EQ:Min}. Similarly, $D$, is passed into Algorithm \ref{ALGO:UQ_JointQuantile} to calculate a two-sided uncertainty interval for the multivariate quantile probability with a desired confidence of $95\%$. The quantile probability for each variate was chosen to be $\tau_{i}=0.90 : i \in \left\{1, \ldots, q \right\}$. Within Algorithm \ref{ALGO:UQ_JointQuantile}, the same 1000 bootstrap resamples were utilized, however, the bootstrap confidence interval method used was altered from the basic percentile to the BCa percentile method. The 95\% confidence interval on $ \tau_{\mathrm{true}}$ is denoted as $ \left[ \tau_{\sm{0.025}},  \tau_{\sm{0.975}} \right]$. Finally, at the end of each Monte Carlo simulation, if the true multivariate coverage was bounded by the uncertainty calculated by Algorithm \ref{ALGO:UQ_JointQuantile}, $\tau_{\sm{0.025}} \le \tau_{\mathrm{true}} \le  \tau_{\sm{0.975}}$, a value of true was returned, else false. The final column of Table \ref{TBL:QuantileUQ} tabulates the calculated confidences for each combination of variables performed in the simulation. Overall, for numerous sample sizes and variates, Algorithm \ref{ALGO:UQ_JointQuantile} provides an overall confidence of $94\%$ which is very close to the true expected confidence value of $95\%$.

\begin{table}[H]
\centering
\begin{tabular}{||c||c||c||} 
\hhline{|t:=:t:=:t:=:t|}
$q$ & $n$                        & $p$     \\ 
\hhline{|:=::=::=:|}
2 & 30                        & 0.93  \\ 
\hhline{|:=::=::=:|}
2 & 300                       & 0.95  \\ 
\hhline{|:=::=::=:|}
2 & 3000                      & 0.89  \\ 
\hhline{|:=::=::=:|}
3 & 30                        & 0.94  \\ 
\hhline{|:=::=::=:|}
3 & 300                       & 0.98  \\ 
\hhline{|:=::=::=:|}
3 & 3000                      & 0.93  \\ 
\hhline{|:=::=::=:|}
4 & 30                        & 0.92  \\ 
\hhline{|:=::=::=:|}
4 & 300                       & 0.95  \\ 
\hhline{|:=::=::=:|}
4 & 3000                      & 0.93  \\ 
\hhline{|:=:b:=::=:|}
\multicolumn{2}{||c||}{AVE}   & 0.94  \\
\hhline{|b:==:b:=:b|}
\end{tabular}
\caption{\label{TBL:QuantileUQ}Numerical simulation results from 100 Monte Carlo simulations for each combination of variables on the joint multivariate quantile probability with a desired confidence of $95\%$. Each individual variate had a quantile probability with the associated value of, $\tau_{i}=0.90 : i \in \left\{1, \ldots, q \right\}$.}
\end{table}

\subsubsection{Algorithm 2 - Confidence Intervals on Multivariate Normal Quantiles}\label{Numerical_Simulation_Algo2}
\begin{figure}[H] 
\includegraphics[width=0.66 \textwidth]{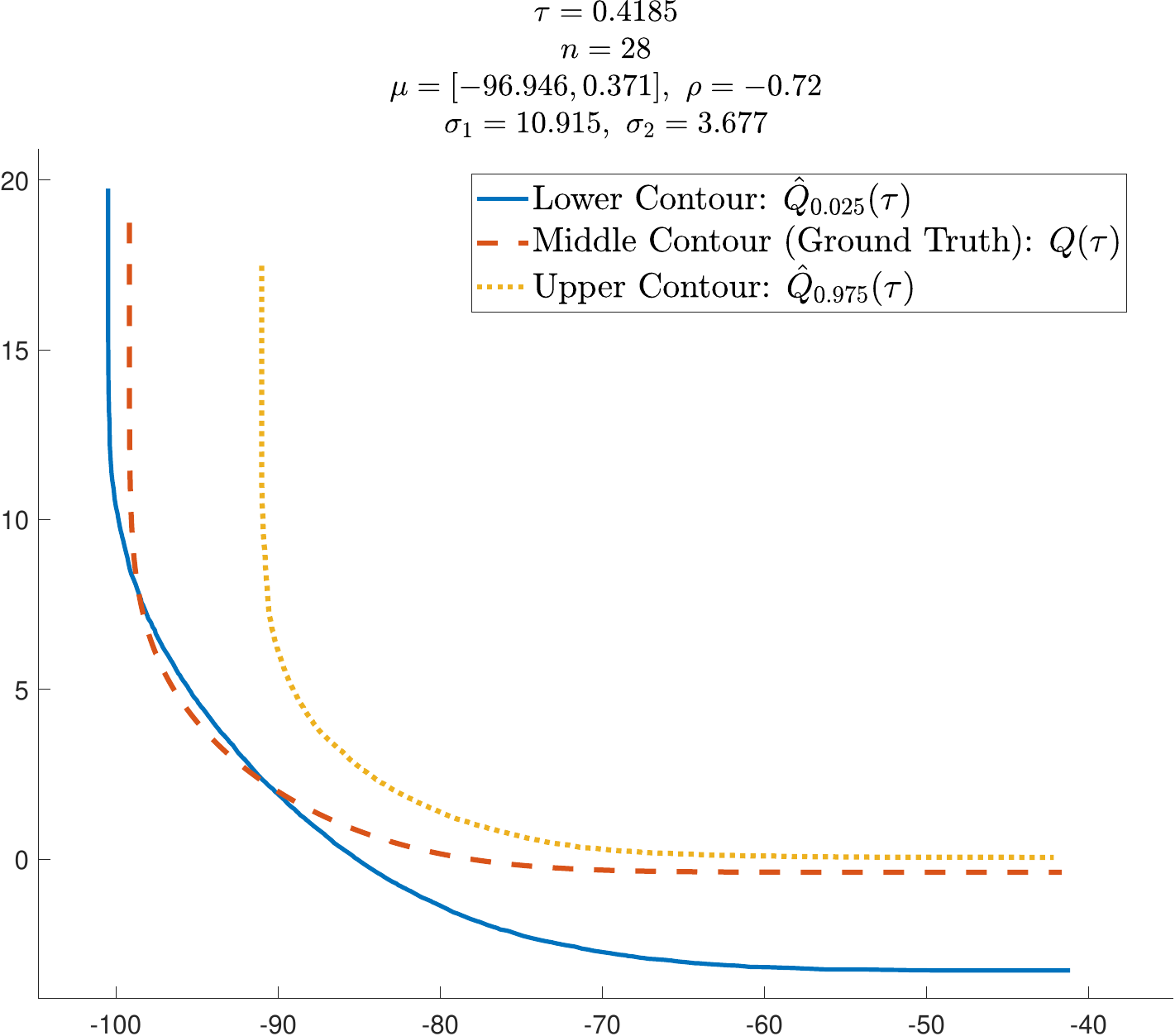}
\centering
\caption{The red dashed contour is the ground truth, $Q(\tau)$, for this bivariate distribution for a quantile probability of $\tau=0.4185$. The solid blue contour, $\hat{Q}_{0.025}(\tau)$, and the dotted yellow contour, $\hat{Q}_{0.975}(\tau)$, denote the lower and upper confidence intervals for the multivariate quantile for an arbitrary sample size of $n=28$. One can see that the true quantile temporarily intersects the blue contour representing the lower confidence bound on the quantile.}
\label{Plot:CDF_Contours_MC_Issues}
\end{figure}

Designing a numerical simulation to test Algorithm \ref{ALGO:CI_MVQuantile} is not a trivial task due to the dimensionality and complexity of the problem space. At this current moment, we believe that the current computation burden prevents running a simulation study directly for confidence intervals on multivariate quantiles for three variates. In lieu of this, the simulation study will use a modified version of Algorithm \ref{ALGO:CI_MVQuantile} designed for two variates. There are numerous combinations of variables that must be tested against to verify that that the methodologies provide the correct solution for a particular confidence value. For bivariate distributions these variables are over the space of all possible correlation coefficients, $-1 \le \rho \le 1$, quantile probabilities, $0 \le \tau \le 1$, and a host of possible sample sizes. Normalization of the data allows one to avoid testing over the space of all possible mean vectors and variances. Figure \ref{Plot:CDF_Contours_MC_Issues} demonstrates the complexity in verifying confidence intervals on CDF based multivariate quantiles. The example given is for a bivariate distribution with arbitrary parameters shown in the plot title. The red dashed contour is the ground truth, $Q(\tau)$, for this bivariate distribution with  $\tau=0.4185$. The solid blue contour, $\hat{Q}_{0.025}(\tau)$, and the dotted yellow contour, $\hat{Q}_{0.975}(\tau)$, denote the lower and upper confidence intervals for a two-sided 95\% confidence interval for the multivariate quantile for an arbitrary sample size of $n=28$. One can see that the true quantile contour temporarily intersects the blue contour representing the lower confidence bound on the quantile contour. Ideally, the lower and upper confidence bounds on the $\tau^{\mathrm{th}}$ quantile would ``sandwich" the ground truth contour. When running Monte Carlo simulation studies for confidence intervals on multivariate quantiles, one could classify any departure from the confidence bounds as false. However, this classification is exceedingly conservative as it foregoes most of the contour. Since the contours are technically an infinite set of points that satisfy \eqref{EQ:MV_Quantile}, any notion of proportion or ``majority" in this context is an ill-posed statement. Therefore, running a simulation in this manner would provide erroneous $p$-values for the presented method.

The authors propose a simulation based on the proportion of the multivariate population that the multivariate quantile bounds; the bivariate case is shown below:
\begin{gather}
\beta = \int \int_{\Omega}  f(x_1, x_2) \ dx_1 \  dx_2 :\nonumber  \\ 
\Omega = \left\{ (x_1, x_2) \  | \ \boldsymbol{x} \le Q(\tau) \right\}.   \label{EQ:NewBeta}
\end{gather}

A bivariate numerical simulation study was performed utilizing $\hat{\beta}$. Three different correlation coefficients, $\rho=\left\{-0.9, 0 , 0.9 \right\}$, and quantile probabilities, $\tau=\left\{0.7, 0.8 , 0.9 \right\}$, were used.  For each combination of the correlation coefficients and quantile probabilities, $\hat{\beta}$, was computed as the ground truth using $Q(\tau)$. The numerical simulation study utilizes 2 different sample sizes, $n=\left\{30, 300 \right\}$ to test the efficacy of the methodology for constructing confidence intervals on CDF based multivariate quantiles. For each combination of $n$, 250 Monte Carlo simulations were performed where a sample size of $n$ was drawn from the true bivariate distribution. A one-sided, upper, confidence interval on the multivariate quantile was performed for that particular sample at a confidence value of 95\%. The upper bound of the confidence interval was utilized to calculate $\hat{\beta}_{i} : i = 1, \ldots, 250$ using the same one hundred thousand random multivariate samples. Finally, after all 250 Monte Carlo simulations were performed, a $p$-value was estimated, $\mathrm{mean}(\hat{\beta}_{i} \ge \hat{\beta})$. This procedure was carried out for all the combinations of variables, shown as the three columns, in Table \ref{TBL:CI_CDFQuantile}. The number of Monte Carlo simulations may seem small with 250, but within each of those simulations, 1000 Monte Carlo bootstraps are performed to construct the contours for the outer Monte Carlo simulation. Therefore, the simulation study presented is very computationally time consuming, taking 2-3 days to compute. Hence, the authors chose to test the bivariate case of Algorithm \ref{ALGO:CI_MVQuantile} which differs only in the tessellation of the support for the distribution.

One can see that the $p$-values calculated are highly correlated with the correlation coefficient, shown as the last column of Table \ref{TBL:CI_CDFQuantile}. When $\rho=-0.9$ the $p$-value is much greater than the theoretical value of $p=0.95$, however, the method is maintaining a conservative estimate. When $\rho \ge 0$, one can see that the $p$-values are close to the expected theoretical value of $p=0.95$. The methodology does not appear to have a bias towards the smaller sample size of $n=30$ compared to when $n=300$ due to using the BCa bootstrap confidence interval method. The results provided here confirm the efficacy methodology at least in the bivariate case. For higher dimension cases, if the support can be tessellated sufficiently, then the method should extrapolate to accommodate them. 

\begin{table}[H]
\centering
\begin{tabular}{|c|c|c|c|c|} 
\hhline{|t:=:t:=:t:=:t:=:t:=:t|}
\multicolumn{1}{||c||}{$\rho$} & \multicolumn{1}{c||}{$\tau$} & \multicolumn{1}{c||}{$n$} & \multicolumn{1}{c||}{$p$} & \multicolumn{1}{c||}{average}  \\ 
\hhline{|b:=:b:=:b:=:b:=:b:=:b|}
\multirow{6}{*}{-0.9}       & \multirow{2}{*}{0.7}      & 30                      & 0.984                   & \multirow{6}{*}{0.997}     \\ 
\cline{3-4}
                            &                           & 300                     & 1                       &                            \\ 
\cline{2-4}
                            & \multirow{2}{*}{0.8}      & 30                      & 1                       &                            \\ 
\cline{3-4}
                            &                           & 300                     & 1                       &                            \\ 
\cline{2-4}
                            & \multirow{2}{*}{0.9}      & 30                      & 1                       &                            \\ 
\cline{3-4}
                            &                           & 300                     & 1                       &                            \\ 
\hline
\multirow{6}{*}{0}          & \multirow{2}{*}{0.7}      & 30                      & 0.960                   & \multirow{6}{*}{0.951}     \\ 
\cline{3-4}
                            &                           & 300                     & 0.956                   &                            \\ 
\cline{2-4}
                            & \multirow{2}{*}{0.8}      & 30                      & 0.940                   &                            \\ 
\cline{3-4}
                            &                           & 300                     & 0.956                   &                            \\ 
\cline{2-4}
                            & \multirow{2}{*}{0.9}      & 30                      & 0.932                   &                            \\ 
\cline{3-4}
                            &                           & 300                     & 0.964                   &                            \\ 
\hline
\multirow{6}{*}{0.9}        & \multirow{2}{*}{0.7}      & 30                      & 0.944                   & \multirow{6}{*}{0.944}     \\ 
\cline{3-4}
                            &                           & 300                     & 0.941                   &                            \\ 
\cline{2-4}
                            & \multirow{2}{*}{0.8}      & 30                      & 0.932                    &                            \\ 
\cline{3-4}
                            &                           & 300                     & 0.964                   &                            \\ 
\cline{2-4}
                            & \multirow{2}{*}{0.9}      & 30                      & 0.948                   &                            \\ 
\cline{3-4}
                            &                           & 300                     & 0.936                    &                            \\ 
\hline
\multicolumn{4}{|c|}{average}                                                                                & 0.964                      \\
\hline
\end{tabular}
\caption{\label{TBL:CI_CDFQuantile}A bivariate numerical simulation study was performed utilizing $\hat{\beta}$. Three different correlation coefficients, $\rho=\left\{-0.9, 0 , 0.9 \right\}$, and quantile probabilities, $\tau=\left\{0.7, 0.8 , 0.9 \right\}$, were used.  For each combination of the correlation coefficients and quantile probabilities, $\hat{\beta}$, was computed as the ground truth using $Q(\tau)$. The numerical simulation study utilizes 2 different sample sizes, $n=\left\{30, 300 \right\}$ to test the efficacy of the methodology for constructing confidence intervals on CDF based bivariate quantiles.}
\end{table}

\subsubsection{Algorithm 3 - Confidence Intervals of Critical Points on Multivariate Quantiles}\label{Numerical_Simulation_Algo3}
Designing a simulation study for Algorithm \ref{ALGO:CI_CriticalPoints} requires a different approach due to the nature of the problem space. There is ambiguity with how to classify whether the true critical point is within the uncertainty interval for the critical points. Ideally, $\hat{\boldsymbol{v}}_{\frac{\alpha}{2}}(\tau) \le \boldsymbol{v}(\tau) \le \hat{\boldsymbol{v}}_{1 - \frac{\alpha}{2}}(\tau)$, but one must pick a metric for measuring this criterion because the critical points are not scalars but vectors. For example, if one element in the uncertainty interval for the critical point vector doesn't satisfy this criteria, it may be too conservative to reject the entire critical point. To avoid having to pick an arbitrary metric, one can utilize the fact that the CDF of the true critical point provides the $\tau^{\mathrm{th}}$ quantile probability, $F(\boldsymbol{v}(\tau))=\tau$. The corollary is that one can now assess the probability whether the CDF value of the one-sided, $p=95\%$, confidence interval on the critical point is at least greater or equal to the $\tau^{\mathrm{th}}$ quantile probability. One can therefore verify that the confidence interval on the critical point is applicable to the correct quantile probability. It is assumed that the confidence interval on the critical point still maintains the property that it is the point associated with the $\tau^{\mathrm{th}}$ quantile probability with highest PDF value. With very large sample sizes, the confidence interval on the critical point does converge to the theoretical critical point. 

Therefore a numerical simulation study was performed for three different number of variates, $q \in \left\{ 2, 3, 4 \right\}$, two different sample sizes, $n \in \left\{30, 300 \right\}$, and three different quantile probabilities, $\tau \in \left\{0.7, 0.8 , 0.9 \right\}$. For each combination of variables, 250 Monte Carlo simulations were performed. For each Monte Carlo simulation, $n$ number of random vectors were drawn from a standardized $q$-variate normal distribution with a random correlation matrix, $C_q$, using the $C$-vine method where $c=2$. For each Monte Carlo simulation, a one-sided, upper, confidence interval on the critical point for the $\tau^{\mathrm{th}}$ multivariate quantile probability was calculated, denoted as, $\boldsymbol{v}_{(1-\alpha, i)}: i = 1, \ldots, 250$, where $i$ denotes the $i^{\mathrm{th}}$ Monte Carlo simulation. The CDF value of the the critical point thus forms a one-sided, upper, confidence bound on the $\tau^{\mathrm{th}}$ quantile probability, denoted as,
\begin{align}
\tau_{(1-\alpha, i)} = F\bigg( \hat{\boldsymbol{v}}_{(1-\alpha, i)} \bigg). \label{EQ:Algo3_a}
\end{align}
Finally, after all 250 Monte Carlo simulations were performed, a $p$-value was estimated using \eqref{EQ:Algo3_a} as,
\begin{gather}
P \Big(\tau_{(1-\alpha, i)} \ge \tau \Big) = \mathrm{mean} \Big(\tau_{(1-\alpha, i)} \ge \tau \Big). \label{EQ:Algo3}
\end{gather}
This procedure was carried out for all the combinations of variables, shown as the three columns, in Table \ref{TBL:CI_CritPoint}. For a one-sided confidence interval with 95\% confidence one expects the $p$-values to be close to 0.95 but the simulation results were 0.984. The default convergence for options for the {\myfont qmvnorm()} is a maximum of 500 iterations or tolerance on the quantile of 0.001. The simulation results provided here increased this default value to a maximum of 5000 iterations which improved the average $p$-value for the entire simulation from 0.993 to 0.984. The results from the numerical simulation in Table \ref{TBL:CI_CritPoint}  provides robust evidence that the numerical methods provided in \Cref{Method2} for constructing confidence intervals on critical points residing on multivariate quantiles are slightly more conservative than their theoretical confidence value of $p=0.95$. 

\begin{table}[H]
\centering
\begin{tabular}{|c|c|c|c|} 
\hhline{|t:=:t:=:t:=:t:=:t|}
\multicolumn{1}{||c||}{$q$} & \multicolumn{1}{c||}{$n$} & \multicolumn{1}{c||}{$\tau$} & \multicolumn{1}{c||}{$p$}  \\ 
\hhline{|b:=:b:=:b:=:b:=:b|}
\multirow{6}{*}{2}        & \multirow{3}{*}{30}     & 0.7                       & 0.980                    \\ 
\cline{3-4}
                          &                         & 0.8                       & 0.964                    \\ 
\cline{3-4}
                          &                         & 0.9                       & 0.948                    \\ 
\cline{2-4}
                          & \multirow{3}{*}{300}    & 0.7                       & 0.980                    \\ 
\cline{3-4}
                          &                         & 0.8                       & 0.996                    \\ 
\cline{3-4}
                          &                         & 0.9                       & 0.996                    \\ 
\hline
\multirow{6}{*}{3}        & \multirow{3}{*}{30}     & 0.7                       & 0.976                    \\ 
\cline{3-4}
                          &                         & 0.8                       & 0.964                    \\ 
\cline{3-4}
                          &                         & 0.9                       & 0.968                    \\ 
\cline{2-4}
                          & \multirow{3}{*}{300}    & 0.7                       & 0.988                    \\ 
\cline{3-4}
                          &                         & 0.8                       & 0.988                    \\ 
\cline{3-4}
                          &                         & 0.9                       & 0.988                    \\ 
\cline{2-4}
\hline
\multirow{6}{*}{4}        & \multirow{3}{*}{30}     & 0.7                       & 0.992                    \\ 
\cline{3-4}
                          &                         & 0.8                       & 0.992                    \\ 
\cline{3-4}
                          &                         & 0.9                       & 0.988                    \\ 
\cline{2-4}
                          & \multirow{3}{*}{300}    & 0.7                       & 1                        \\ 
\cline{3-4}
                          &                         & 0.8                       & 1                        \\ 
\cline{3-4}
                          &                         & 0.9                       & 1                        \\ 
\hline
\multicolumn{3}{|c|}{average}                                                   & 0.984                    \\
\hline
\end{tabular}
\caption{\label{TBL:CI_CritPoint}A numerical simulation study was performed for three different variates, $q \in \left\{ 2, 3, 4 \right\}$, two different sample sizes, $n \in \left\{30, 300 \right\}$, and three different quantile probabilities, $\tau \in \left\{0.7, 0.8 , 0.9 \right\}$. 250 Monte Carlo simulations were performed generating a $p$-value for each combination of variables. A confidence value of 95\% was used throughout the study.}
\end{table}

\subsection{Shock Data Case Study for Environmental Specification}\label{Shock_Data}
Over the past decade, there has been strong research interest in multi-axis testing as opposed to single-axis testing within the field of mechanical shock and vibration. Yet, many test standards still currently utilize only single-axis shock and vibration testing such as NAVMAT P-9492 from the U.S. Navy or the electronics industry’s JESD22-B103B \citep{noauthor_tmil-std-810_nodate, noauthor_jesd22s_nodate, noauthor_navmat_nodate}. The benefits of multi-axis testing are that it excites all modes simultaneously and induces a more realistic stress loading condition while alleviating the issue where the order in which orthogonal uni-axial excitation is applied has a significant effect on fatigue failure \citep{whiteman_fatigue_2002}. \cite{sisemore_multi-axis_2020} provides information about the development of multi-axis shock test specifications while brashly stating, ``writing a three-axis shock test specification is not particularly complex''. This statement may be true when one knows the true underlying shock excitation or its respective shock response spectra. In practice, one rarely knows the exact environmental vibration or shock excitation levels. \cite{nelson_systematic_2017} provides four different derivation methods to develop six degree of freedom (6-DOF) input specifications for multi-input, multi-output (MIMO) mechanical vibration testing. The current literature neglects to mention about statistical conservatism and how it applies to multi-axis shock and vibration testing. 

Environmental specification implies deriving what are the extreme shock or vibration environmental levels that a system or component may experience in the field in a statistical sense given some specified risk tolerance. For single-axis shock and vibration testing this amounts to using the collected time series acceleration field data from tri-axial accelerometers and transforming the data into the spectral domain; power spectral density (PSD) or shock response spectrum (SRS) curves for shock and vibration data respectively \citep{lalanne_mechanical_2014}. At each frequency value in the spectral domain, the data are assumed to be either normally or log-normally distributed. The Shapiro-Wilk or Anderson Darling test can be used to test for normality. Most engineers who are qualifying a system or component in the domain of shock and vibration are concerned about the extreme responses across a band/range of frequencies, thus a one-sided, upper, normal tolerance bound is often constructed to bound at least a proportion of the data for a particular confidence value. The case study provide here will demonstrate how to derive a three-axis environmental specification for mechanical shock from field data that is statistically conservative. A fully defined 6-DOF vibration test is a $6\times6$ spectral density matrix, the diagonal includes the three translation directions X, Y, and Z and the rotations associated for each spatial direction \citep{jacobs-omalley_methodology_2014, nelson_comparison_2015}. For simplicity, the environmental specification here will be designed for 3-DOF shock test for just the translation directions. 

\begin{figure}[H] 
\includegraphics[width=\textwidth]{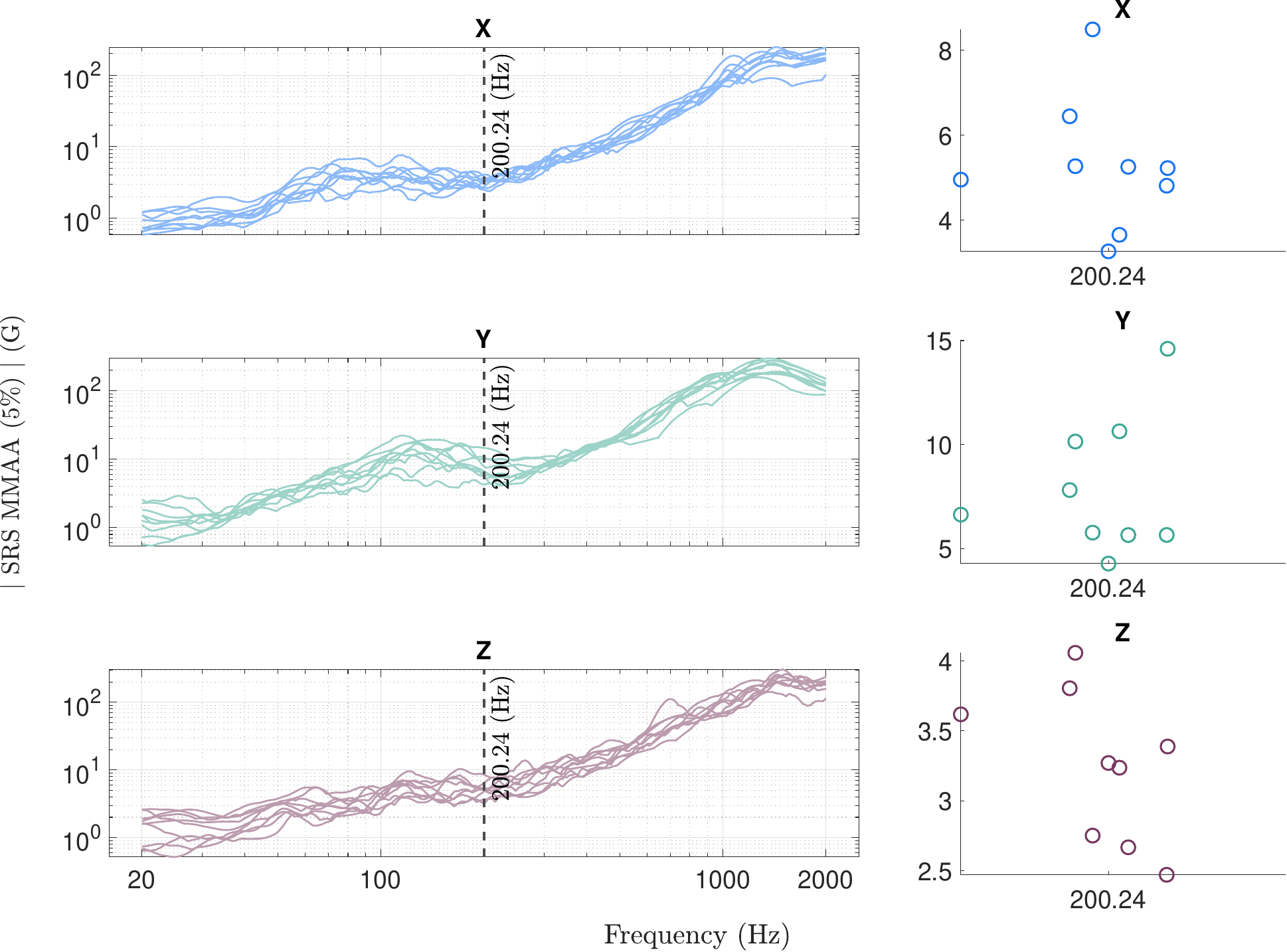}
\centering
\caption{Tri-axial shock data in the shock response spectrum (SRS) domain using the maxi-max absolute acceleration (MMAA) method with a 5\% damping factor. Each row corresponds the SRS MMAA for each spatial direction X, Y, and Z respectively. The first column represents the SRS over the entire frequency domain of interest from 20 Hz to 2000 Hz. The second column are swarm plots of the SRS values at a frequency of 200.24 Hz with random horizontal jitter to avoid overlapping data points for visualization purposes. The nine SRS values for each channel direction will comprise the columns of the data set, $D \in \mathbb{R}^{ (9 \mathrm{x} 3)}$.}
\label{Plot:ShockData3D}
\end{figure}

Environmental field data, time series acceleration, was collected from a test body instrumented with a tri-axial accelerometer. The test body was exposed to the same environmental shock nine times, thus creating an ensemble of tri-axial accelerometer data. Each time series was then transformed into the spectral domain forming the shock data set, SRS in the case for mechanical shock. The SRS calculation utilized the maxi-max absolute acceleration (MMAA) method with a 5\% damping factor \citep{lalanne_mechanical_2014}. The first column of Figure \ref{Plot:ShockData3D} plots the SRS for an ensemble with each curve representing a SRS from each sample. The rows indicate the SRS ensemble for each spatial direction. At each frequency in the SRS domain, one would compute a pointwise, three-axis, environmental specification for mechanical shock. To illustrate this pointwise procedure, the SRS data occurring at an arbitrary frequency of 200.24 Hz (shown as a dashed line for all the plots in the first column) will be further analyzed. The second column of Figure \ref{Plot:ShockData3D} are swarm plots representing the SRS values at a frequency of 200.24 Hz; each row still corresponds the SRS values in the X, Y, or Z spatial directions respectively. 

\begin{table}[H]
\centering
\begin{tabular}{||c||c||c||c||} 
\hhline{|t:=:t:=:t:=:t=:t|}
X & Y & Z & $d_j$     \\ 
\hhline{|:=::=::=:=:t|}
8.49 & 5.76  & 2.75 & 4.99  \\ 
\hhline{|:=::=::=:=:t|}
6.44 & 7.81  & 3.80 & 2.12  \\ 
\hhline{|:=::=::=:=:t|}
5.26 & 5.65  & 2.67 & 1.25  \\ 
\hhline{|:=::=::=:=:t|}
3.27 & 4.27  & 3.27 & 3.36  \\ 
\hhline{|:=::=::=:=:t|}
4.81 & 5.65  & 2.47 &2.33  \\ 
\hhline{|:=::=::=:=:t|}
3.66 & 10.64  & 3.24 & 1.95  \\ 
\hhline{|:=::=::=:=:t|}
4.96 & 6.63  & 3.62 & 1.07  \\ 
\hhline{|:=::=::=:=:t|}
5.23 & 14.61  & 3.39 & 4.64  \\ 
\hhline{|:=::=::=:=:t|}
5.27 & 10.14  & 4.06 & 2.28  \\ 
\hhline{|:=::=::=:=:t|}
\end{tabular}
\caption{\label{TBL:ExampleData} The first three columns of the table are the tabulated data from the second column of the swarm plots in Figure \ref{Plot:ShockData3D} for the three spatial directions X, Y and Z respectively. Each row of the first three columns is considered as one multivariate observation. The final column of the table is the squared Mahalanobis distance, $d_j$, for each multivariate observation and is calculated using \eqref{EQ:Mahalanobis} for the purpose of checking the multivariate normality assumption.}
\end{table}

\begin{figure}[H] 
\includegraphics[width= 0.6666 \textwidth]{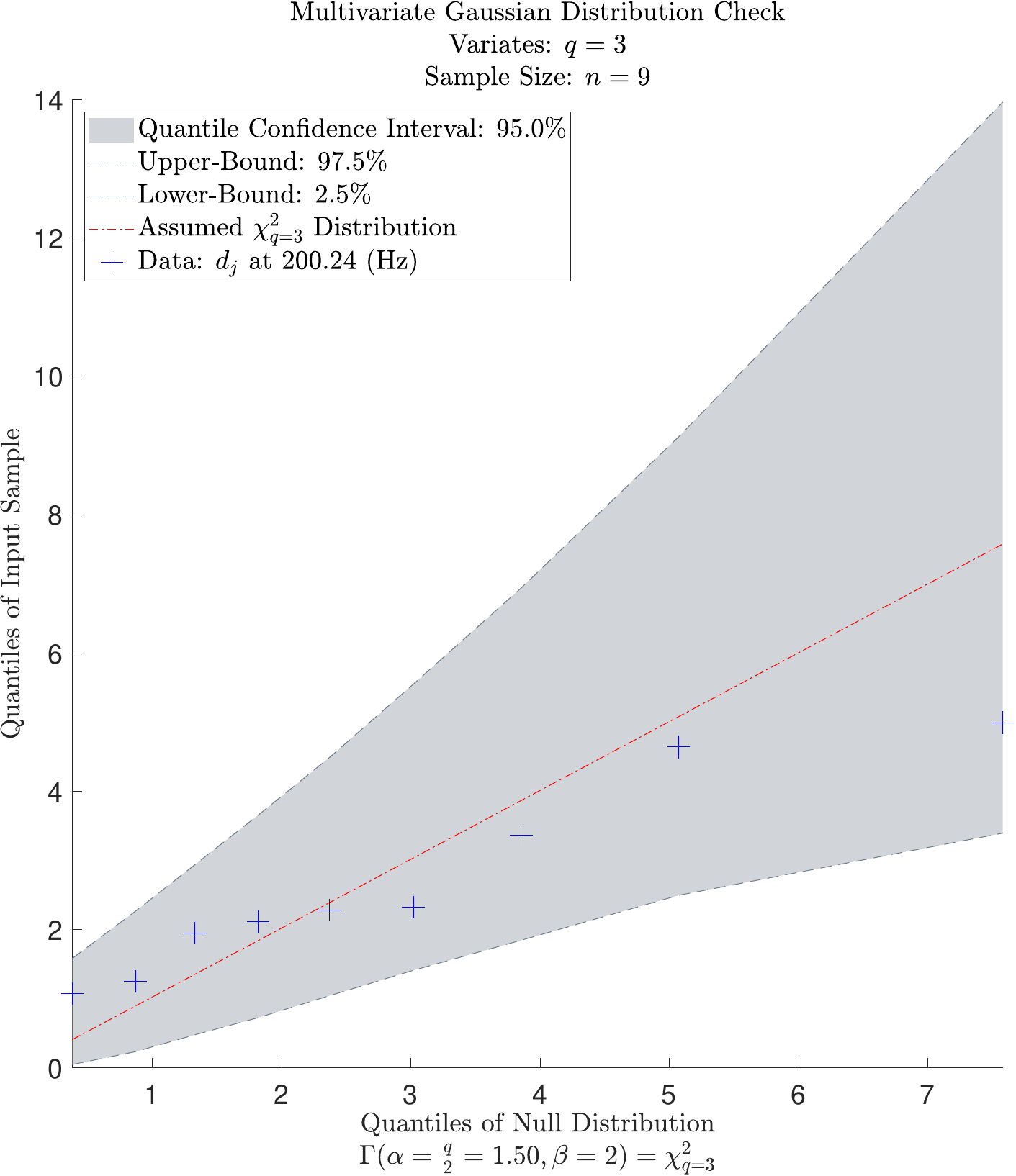}
\centering
\caption{QQ plot for checking the multivariate Gaussian assumption. The grey shaded region represents a 95\% confidence region with a sample size of $n=9$ for the squared Mahalanobis distance for trivariate normal data. The squared Mahalanobis distance follows a chi-square distribution with $q$-degrees of freedom, three degrees of freedom in this case, $\chi_{q=3}^2$. The dashed grey lines represent the lower and upper confidence bounds of the region representing 2.5\% and 97.5\% confidence respectively. The red dash-dotted line represents the theoretical QQ line for $\chi_{q=3}^2$. The blue crosses represent the squared Mahalanobis distance for each sample, $d_j$, from the multivariate data tabulated in Table \ref{TBL:ExampleData}.}
\label{Plot:Shock3D_QQ}
\end{figure}

For convenience, the SRS values occurring at 200.24 Hz for each spatial direction are also tabulated as the first three columns in Table \ref{TBL:ExampleData}. Nine SRS values for each of the three spatial directions compose the columns of the data set, $D \in \mathbb{R}^{ (9 \mathrm{x} 3)}$. The numbers in the last column of Table \ref{TBL:ExampleData} are the squared Mahalanobis distances, $d_j$, associated with each trivariate sample, calculated using \eqref{EQ:Mahalanobis}. Figure \ref{Plot:Shock3D_QQ} provides a quantile-quantile (QQ) plot for the data set as a graphical method to test for multivariate normality. The grey shaded region represents a 95\% confidence region with a sample size of $n=9$ for the squared Mahalanobis distance for trivariate normal data. The squared Mahalanobis distance approximately follows a chi-square distribution with $q$-degrees of freedom, three degrees of freedom in this case, $\chi_{q=3}^2$ \citep{johnson_applied_2007}. In Addition to the QQ plot, two statistical hypothesis tests were used to check the assumption of multivariate normally distributed data. An Anderson-Darling and a Kolmogorov-Smirnov test were used to assess if sample data, $d_j$, can be assumed to approximately originate from a population with a $\chi_{q=3}^2$ distribution using a significance value of $\alpha=0.05$. The p-values for the test were 0.6054 and 0.7185 for the Anderson-Darling and a Kolmogorov-Smirnov test respectively.  Since all the squared Mahalanobis distances from the data set reside within these confidence bounds, and the p-values were all greater than the significance value of $\alpha=0.05$, one can assume that the data set can be adequately represented as trivariate normally distributed.  

\begin{sidewaysfigure}[h] 
\includegraphics[width=\textwidth]{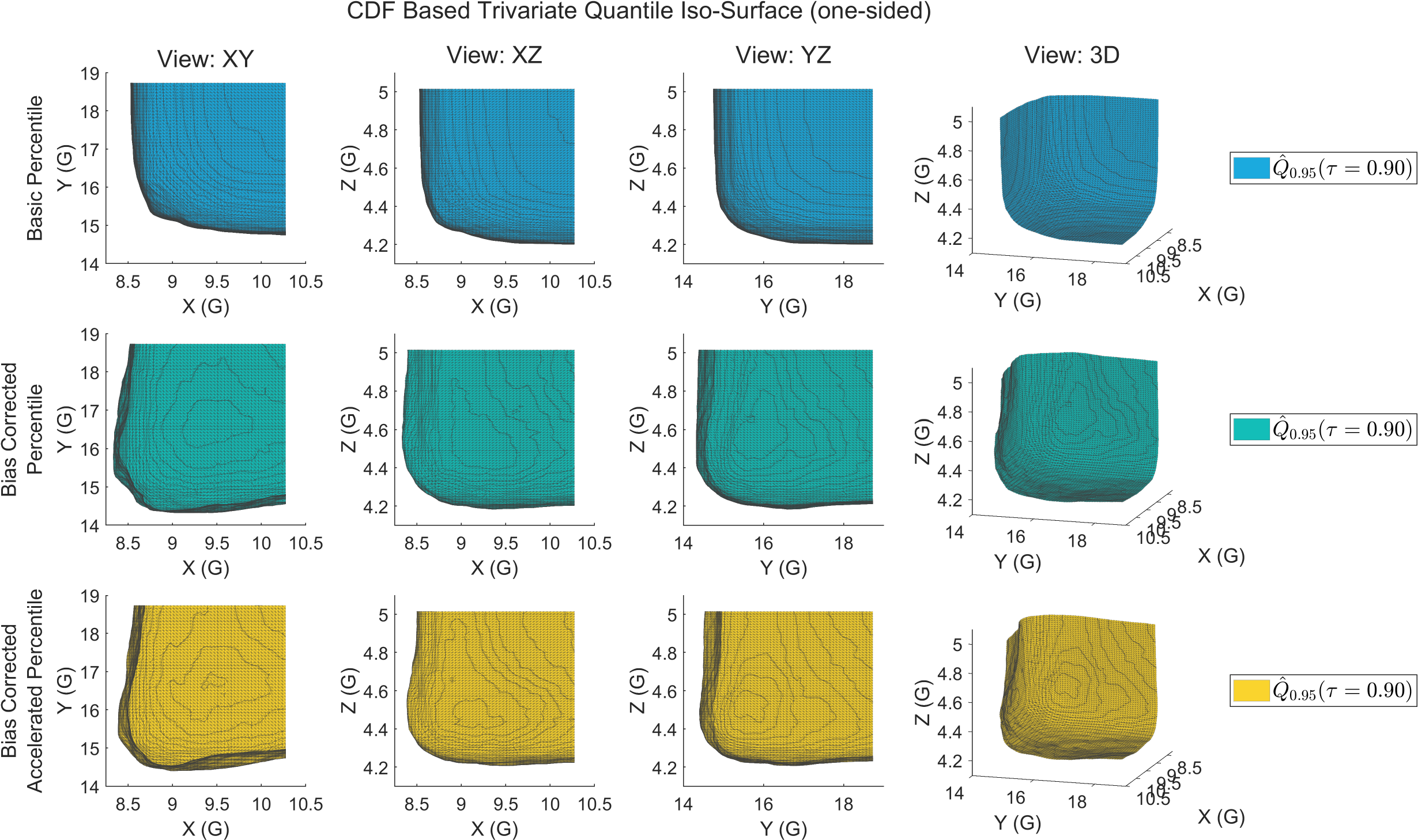}
\centering
\caption{3D plot of the uncertainty on a one-sided, CDF based, trivariate quantile iso-surface with a confidence value of 95\%. The first three columns depict the different views which are normal to the plane corresponding to the XY, XZ, and YZ axis. The fourth column provides a 3D perspective of the iso-surface. The three rows depict the various bootstrap confidence interval methods: basic percentile, bias corrected percentile, and bias corrected and accelerated (BCa) percentile.} 
\label{Plot:ShockData3D_IsoSurfaces}
\end{sidewaysfigure}

Figure \ref{Plot:ShockData3D_IsoSurfaces} is a 3D plot of the uncertainty on a one-sided, CDF based, trivariate quantile iso-surface with a confidence value of 95\% using the data in Table \ref{TBL:ExampleData}. The first three columns depict the different views which are normal to the plane corresponding to the XY, XZ, and YZ axis. The fourth column provides a 3D perspective of the iso-surface. The three rows depict the various bootstrap confidence interval methods: basic percentile, bias corrected percentile, and bias corrected and accelerated (BCa) percentile. A total number of $b=2000$ bootstrap resamples were performed. Significantly more distortion of the iso-surface incur when utilizing the BCa bootstrap method compared to the basic percentile method. These distortions appear as ``waviness" as opposed to smoothness in the iso-surfaces. This result is counter intuitive as the BCa method has precedence of being the more accurate bootstrap confidence interval method. For the bivariate case the BCa  is the preferred method as it maintains smoothness of the quantile contour while correcting for bias and skewness. It is not clear whether this distortion is resolved by increasing the bootstrap sample size, decreasing the mesh tessellation size, or just limitations of the BCa method for bootstrap confidence intervals in higher dimensions. Calculating $\hat{Q}_{0.95}(\tau=0.90)$ requires a significant amount of computation time in $\mathbb{R}^3$ compared to directly calculating the critical points on $Q(\tau)$. The hat symbol on $\hat{Q}_{\gamma} $ denotes that this is an estimated subset of $Q(\tau)$, since $Q(\tau)$ is an infinite set of vectors.

\afterpage{\clearpage}

\begin{figure}[H] 
\includegraphics[width= 0.6666 \textwidth]{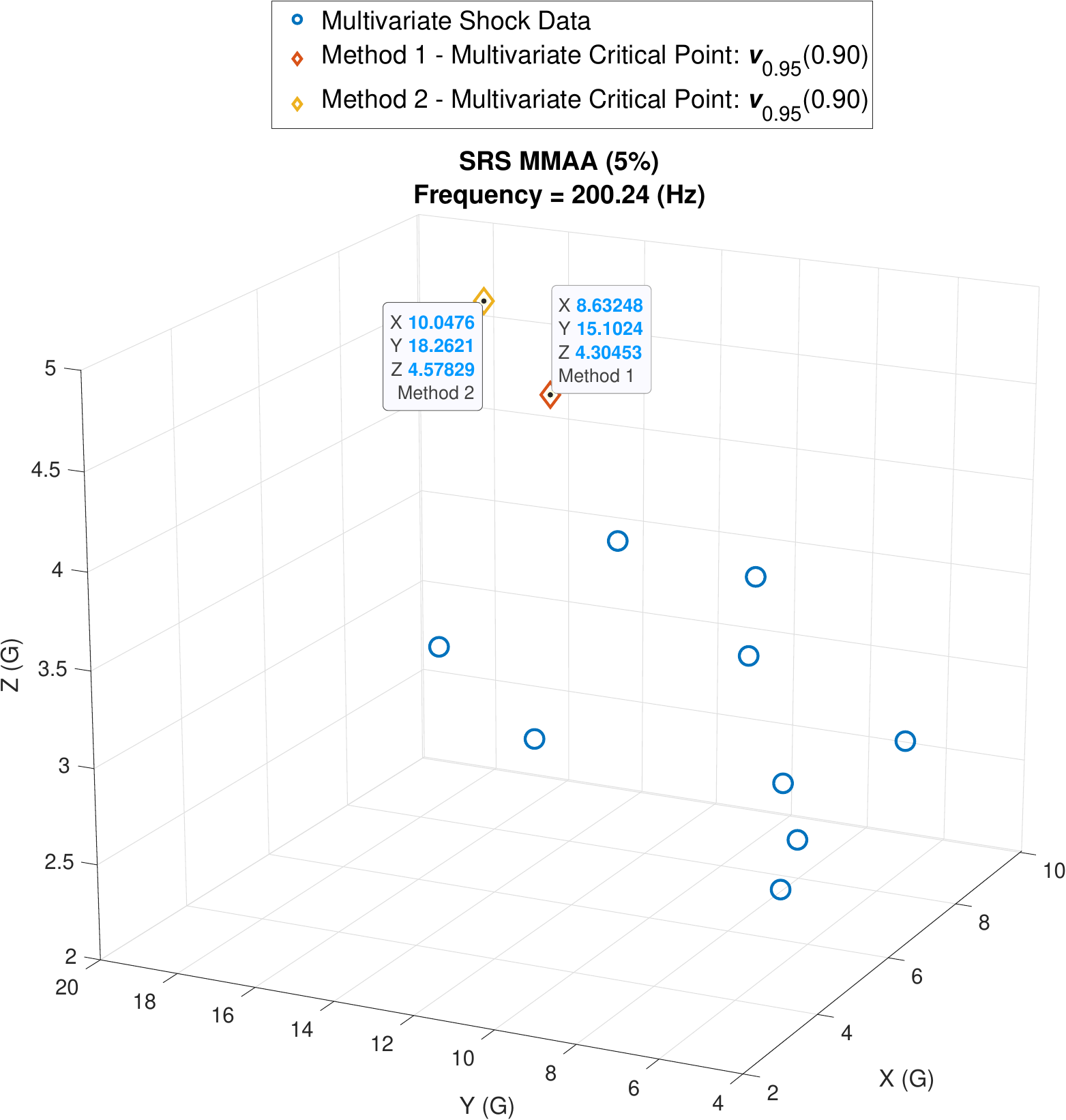}
\centering
\caption{3D scatter plot of the trivariate data found in Table \ref{TBL:ExampleData} along with the one-sided confidence bound on the critical points using method 1 from \cref{Method1} and method 2 from  \cref{Method2}. Method 1 is the critical point from the iso-surface and Method 2 is the direct numerical solution from Method 2, both utilizing a BCa bootstrap confidence interval method.} 
\label{Plot:Shock3D_CriticalPoint}
\end{figure}

Figure \ref{Plot:Shock3D_CriticalPoint} is a 3D scatter plot of the trivariate data found in Table \ref{TBL:ExampleData} along with the critical points using method 1 from \cref{Method1} and method 2 from  \cref{Method2} above. Method 1 is the critical point from the iso-surface using the BCa bootstrap confidence interval on the tessellated CDF. Method 2 is the direct numerical solution also using the BCa bootstrap confidence interval method. Both methods used $b=2000$ bootstrap resamples using a nonparametric bootstrap resampling scheme. The critical point, $\boldsymbol{v}_{0.95}(0.90)$, from method 2 is strictly greater than the critical point from method 1, thus a more conservative value. These results here agree with the numerical simulation study as Method 2 was more conservative than Method 1 as the desired $p$-value was greater than the theoretical value of 0.95 for a confidence level of 95\%. However, the difference between the two critical points is significant. The authors have found that errors start to become significant when calculating the CDF based quantiles for trivariate distributions with a coarse tessellation mesh size of 0.1
Errors begin to occur as a result of estimating the iso-surface from the coarse tessellation of the support. Secondly, estimating the critical point from this coarse iso-surface leads to additional errors. All the errors accrued in the standardized domain are thus scaled when transforming back the original domain of the data. For standard deviations greater than unity, this implies that the errors are multiplied becoming even greater. The difference in critical point values from the various bootstrap confidence interval methods were less significant with errors only in the first decimal place. Therefore, the errors are predominately associated from the differences in the methodologies. If the critical points are only needed, method 2 provides the most accurate solution while being over two orders of magnitude faster (approximately 664x). 

For multi-axis environmental specification in $\mathbb{R}^3$, it is desirable to provide a single conservative point that bounds the majority of the population with the highest probability density. Thus, at a frequency of 200.24(Hz), a 95\% one-sided confidence interval on the critical point of the multivariate quantile $Q(\tau=0.90)$ with the highest PDF value  is $\boldsymbol{v}_{0.95}(0.90) = [10.0476, \    18.2621, \   4.5783] (\text{G})$ using method 2, see Figure \ref{Plot:Shock3D_CriticalPoint}. Therefore, this critical value bounds the extreme responses for a proportion of at least 0.90 from a trivariate population with a given confidence of 95\%. The procedure listed above would be repeated for all subsequent frequencies from the SRS. Coalescing  the values from each confidence interval over all frequencies are then use to construct the final multi-axis shock environmental specification in $\mathbb{R}^3$.

Table \ref{TBL:MV_vs_UNI} provides tabulated results comparing the methodology for confidence intervals of critical points on multivariate quantiles to univariate tolerance intervals for each of the three variates represented as three spatial directions. One can see that the multivariate statistical interval is strictly greater than each of the univariate tolerance intervals which is necessitated by the fact that the confidence interval on the multivariate quantile accounts for each variate but also the correlations between the variates. Utilizing Algorithm \ref{ALGO:UQ_JointQuantile} and a one-sided BCa bootstrap confidence interval, one finds that the uncertainty on multivariate quantile when using concurrent univariate tolerance intervals is $\tau_{J}=0.76302$ with 95\% confidence. As a consequence of neglecting correlation between the variates, the actual joint or multivariate quantile probability is less than the desired quantile probability of $\tau=0.90$. The multivariate methods presented here do not have these issues as they  account for correlations between the variates. The bounds for the X and Y variates using the Bonferroni interval are less than the multivariate method. It is not clear that the Bonferroni method would be strictly less than the multivariate method. The simultaneous tolerance interval $(\UT{T}_{\mathrm{simul}, \beta=0.90})$ does provide a coverage, $\beta=\tau$, with 95\% confidence but $F(\UT{T}_{\mathrm{simul}, \beta=0.90}) \ne \tau$. The multivariate method on the other hand provides a coverage, $\beta \ge \tau$ with $F(\boldsymbol{v}_{0.95}(0.90)) = \tau$ with 95\% confidence.
\begin{table}[H]
\centering
\begin{tabular}{||c||c||c||c||} 
\hhline{|t:=:t:=:t:=:t:=:t|}
\textbf{Statistical
  Interval Type}~                                                                                       & \textbf{X (G)}   & \textbf{Y (G)}   & \textbf{Z (G)}   \\ 
\hhline{|:=::=::=::=:|}
\begin{tabular}[c]{@{}c@{}}Confidence Interval on Critical Point\\of Multivariate Quantile (Method 2)\end{tabular} & 10.0476 & 18.2621 & 4.5783  \\ 
\hhline{|:=::=::=::=:|}
Univariate Tolerance Interval                                                                                      & 9.0081  & 15.9969 & 4.5694  \\ 
\hhline{|:=::=::=::=:|}
\begin{tabular}[c]{@{}c@{}}Bonferroni Simultaneous \\Tolerance Interval\end{tabular}                                & 9.8128  & 17.7368 & 4.8529  \\
\hhline{|b:=:b:=:b:=:b:=:b|}
\end{tabular}
\caption{\label{TBL:MV_vs_UNI} Comparison of the proposed multivariate critical point quantile confidence intervals, univariate tolerance intervals, and the Bonferroni simultaneous univariate tolerance interval. All are one-sided statistical intervals using a confidence value of 95\%  with either a multivariate or univariate quantile probability value of $\tau=0.90$}
\end{table}

 \section{Future Work and Conclusion}\label{Conclusion}
Evaluating a multivariate CDF over a tessellated space in $\mathbb{R}^3$ can take several minutes or longer for a single Monte Carlo simulation which strongly inhibits numerous simulations that are required for constructing bootstrap confidence intervals. The problem exponentially gets worse with further increases in dimensionality. One possibility to alleviate the problem with computing the CDF over a high dimensional space is to empirically calculate the CDF. \cite{langrene_fast_2021} have shown that the multivariate empirical CDF calculation can be fast with a high degree of accuracy. Future work should also consider the use of adaptive mesh refinement after a coarse initial estimates of the CDF based multivariate normal quantile has been calculated. Using an iterative method to refine where the CDF based multivariate normal quantile are located within $\mathbb{R}^q$ will provide results with higher accuracy, minimizing unnecessary CDF calculations over areas where the quantile are not located and minimizing memory usage. Ultimately, CDF based quantile methods consider the tails of the multivariate distributions while elliptical quantiles from traditional tolerance regions do not. 

Three numerical simulation studies have been provided with all of them demonstrating great efficacy. The first numerical study was on quantifying the uncertainty on the multivariate quantile probability, $\tau$, when using concurrent one-sided univariate quantile probabilities. The second numerical study was on the constructing confidence internals on CDF based bivariate normal quantiles, $\hat{Q}_{\gamma}(\tau)$. The third numerical study was on constructing confidence internals on critical points on CDF based multivariate normal quantiles, $\hat{\boldsymbol{v}}_{\gamma}(\tau)$, for $q \in \left\{ 2, 3, 4 \right\}$. The case study provided here demonstrates a beneficial workflow example for utilizing the presented methodologies and the current limitations of them. We have shown the inadequacy of univariate statistical intervals when working with multivariate data. Multi-axis shock and vibration testing provides a domain where multivariate quantiles are salient to properly address conservatism when developing environmental test specifications. The methodologies and algorithms presented here will also help practitioners using multivariate data with techniques to address the concept of a ``one-sided" statistical interval for tolerance and quantile applications in a multivariate space.

\begin{appendix}

\section{Bounds on the Multivariate Quantile Probability}\label{appendix:a}
 The easiest bound for the multivariate quantile probability is the trivial solution when all the correlations between the variates are 1, i.e., $1 = \rho_{\sm{XY}} = \rho_{\sm{XZ}} = \rho_{\sm{YZ}} $. For this trivial case, since all the variates are co-linear, the multivariate quantile probability, $\tau_{\sm{XYZ}}$, is identical to the individual univariate quantile probability values,
\begin{align}
\tau_{\sm{XYZ}} = \tau_{i} : i \in \left\{X, Y, Z\right\}.
\label{EQ:Coliner}
\end{align}
A lower bound on the multivariate quantile probability for all three variates can be calculated using the Bonferroni inequality provided by \cite{meeker_statistical_2023}, Section D.7.3,
\begin{align}
\tau_{\sm{XYZ}} \geq  1 - \left(   \sum_{i=\left\{X, Y, Z\right\}} (1-\tau_{i}) \right) ,
\label{EQ:Bonferroni}
\end{align}
which occurs when all the variates are maximally, negatively, correlated: $-1 = \rho_{\sm{XY}} = \rho_{\sm{XZ}} = \rho_{\sm{YZ}}$.
In the case where all the variates are uncorrelated, i.e., $0 = \rho_{\sm{XY}} = \rho_{\sm{XZ}} = \rho_{\sm{YZ}}$, the multivariate quantile probability can be calculated as, 
\begin{align}
\tau_{\sm{XYZ}} =   \prod_{i=\left\{X, Y, Z\right\}} \tau_{i} .
\label{EQ:Uncorrelated}
\end{align}

\section{Calculating the Squared Mahalanobis Distances}\label{appendix:b}
The proportion of samples that reside inside the tolerance region can be determined by using the squared Mahalanobis distance. Calculating the squared Mahalanobis distances, $d_j$, for the $j^{\mathrm{th}}$ multivariate sample, $\boldsymbol{x}_j$, can be performed as such: 

\begin{align}
   d_j = h(\boldsymbol{x}_j, \bar{\boldsymbol{x}}, S) := (\boldsymbol{x}_j- \bar{\boldsymbol{x}}) S^{-1} (\boldsymbol{x}_j - \bar{\boldsymbol{x}})^T , \label{EQ:Mahalanobis}
\end{align}
where $\bar{\boldsymbol{x}}$ denotes the sample mean vector, $S$ denotes the sample covariance matrix, and the superscript $T$ denotes the transpose operator. Both $\bar{\boldsymbol{x}}$ and $S$ are plugin estimates for $\boldsymbol{\mu}$ and $\Sigma$, respectively.  One can also calculate the squared Mahalanobis distances for all the multivariate samples in the data:
\begin{align}
   \boldsymbol{d} = h(D, \bar{\boldsymbol{x}}, S) := \mathrm{diag}\bigg( (D - \bar{\boldsymbol{x}}) S^{-1} (D -\bar{\boldsymbol{x}})^T \bigg), 
\end{align}
where $\boldsymbol{d}$ is a vector of all the squared Mahalanobis distances for every multivariate sample. Therefore the $j^{\mathrm{th}}$ element in $\boldsymbol{d}$ is associated to the squared Mahalanobis distance for the $j^{\mathrm{th}}$ multivariate sample in $D$.

\section{Properties of CDF Based  Quantile Contours, Iso-Surfaces, and Hyper Iso-Surfaces} \label{appendix:c}
When the multivariate distribution's parameters are known, one can numerically estimate the multivariate quantiles by first meshing the support for the distribution, commonly done through the tessellation of a $q$-dimensional Euclidean space. The multivariate CDF can then be calculated for any arbitrary distribution over the tessellated space using most modern computational languages such as MATLAB or R. For bivariate and trivariate distributions, adaptive quadrature based on the methods developed by \cite{drezner_computation_1994}, \cite{drezner_computation_1990}, and \cite{genz_numerical_2004} are often employed. However, when there are four or more dimensions, a quasi-Monte Carlo integration algorithm based on methods developed by \cite{genz_numerical_1999} and \cite{genz_comparison_2002} are often used. Most multivariate CDF functions also provide an option to control the absolute error tolerance, typically on the order of 1e-4 or less. After the CDF values are calculated over the space, one can then extract locations over the support where approximately equal CDF values are located, thus creating the CDF based quantile contours for the bivariate case and CDF based quantile iso-surfaces in the trivariate case. 

\begin{sidewaysfigure}[h!] 
\includegraphics[width=\textwidth]{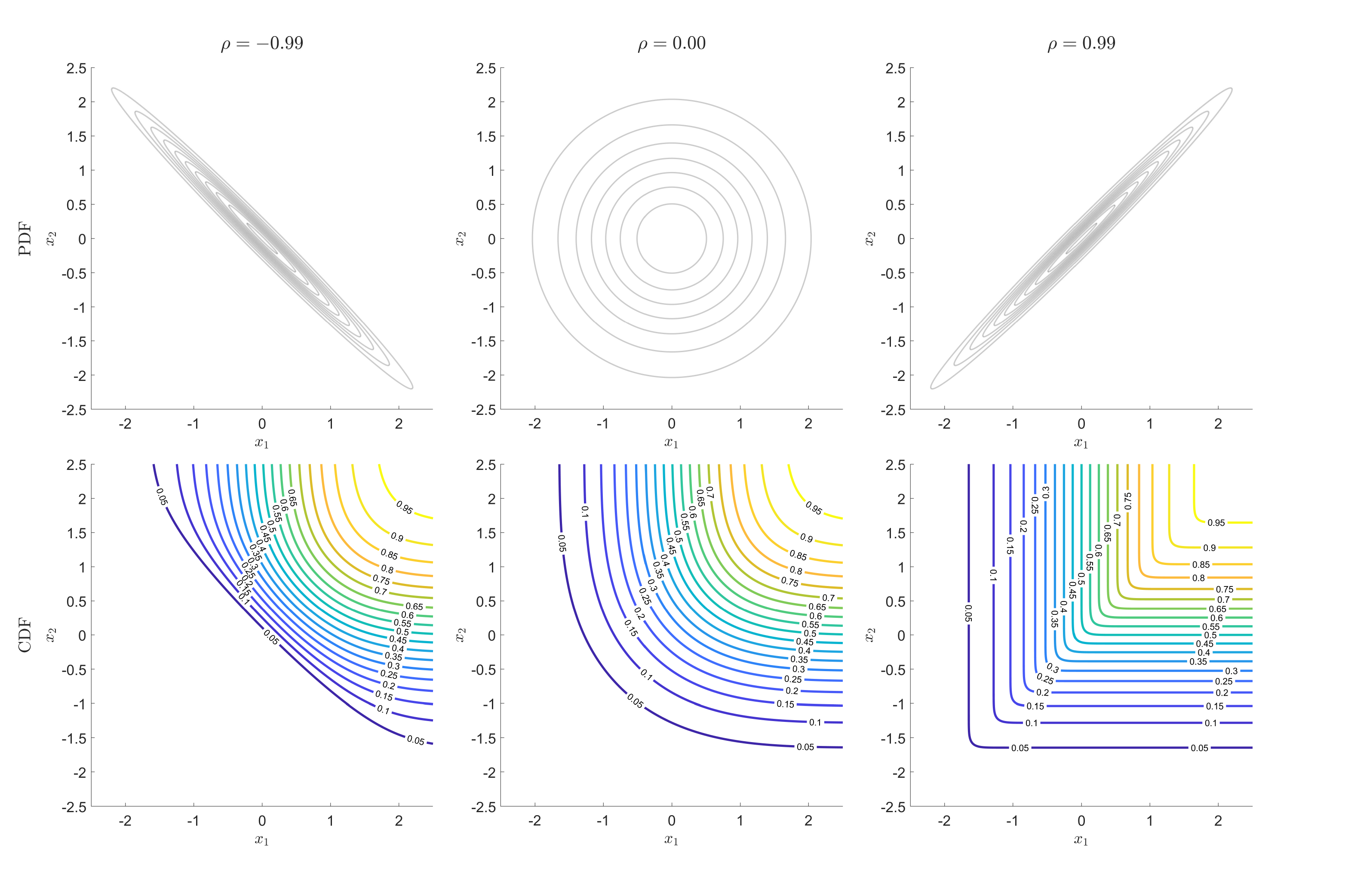}
\centering
\caption{CDF Based bivariate normal quantiles from $\tau = \left\{0.05, 0.1, \ldots, 0.90, 0.95 \right\}$, shown as a gradient of colored contours. At low quantile probability values, such as $\tau=0.05$, there exists strong differences between the quantile contours for various correlation coefficients. At larger quantile probability values, such as $\tau=0.95$, the differences between the quantiles from various correlation coefficients become harder to identify.}
\label{Plot:Bivariate_PDF_CDF}
\end{sidewaysfigure}

\begin{sidewaysfigure}[h!]
\centering
\subfigure[CDF based quantile contour at a low quantile probability value, $\tau = 0.01$, shown in purple.]{%
\resizebox*{0.45 \textwidth}{!}{\includegraphics{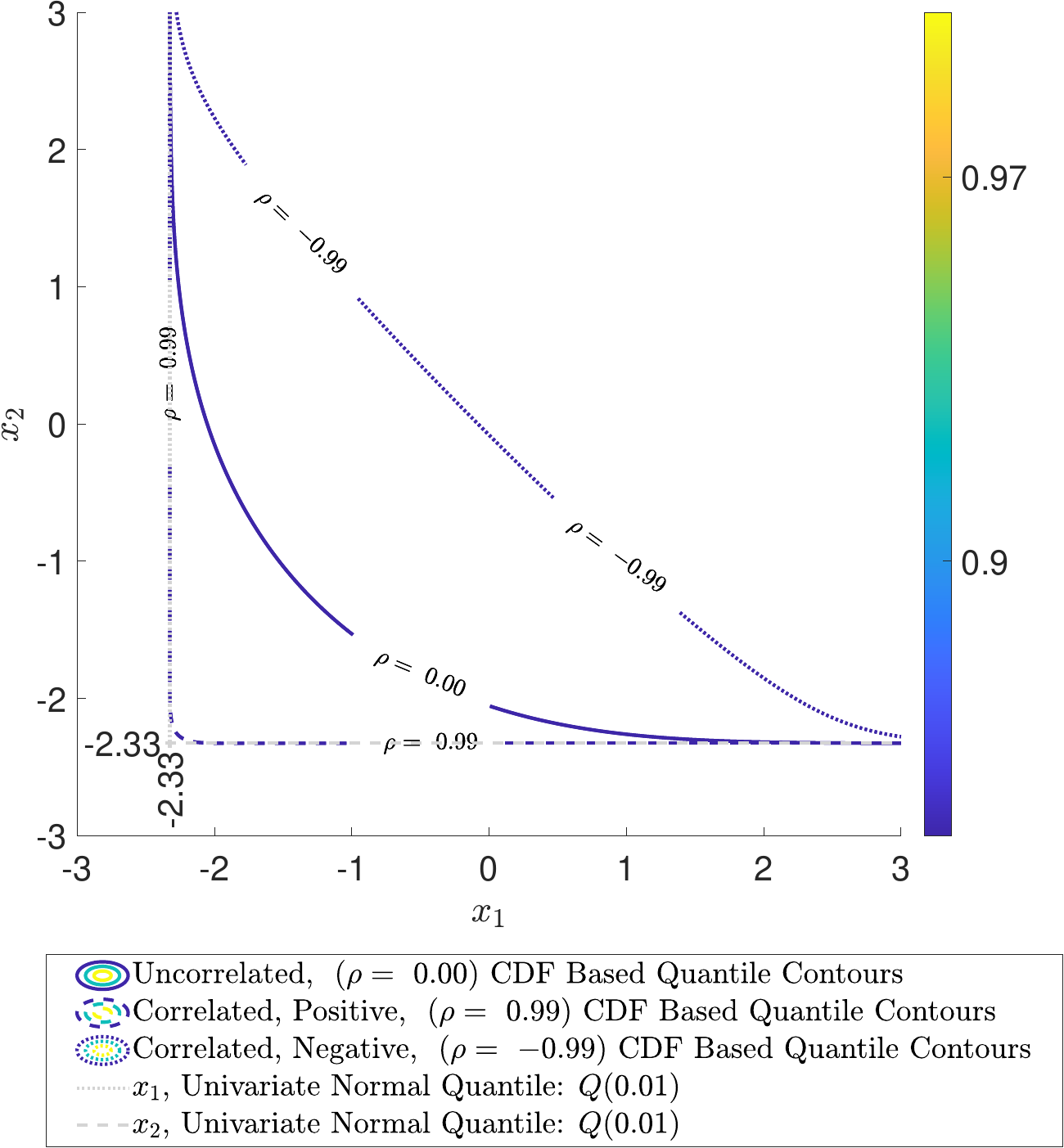}} \label{Plot:Bivariate_CDF_A}}
\subfigure[CDF based quantile contours at two different large quantile probability values, $\tau = \left\{0.90 , 0.97 \right\} $, shown in blue and gold respectively.]{%
\resizebox*{0.45 \textwidth}{!}{\includegraphics{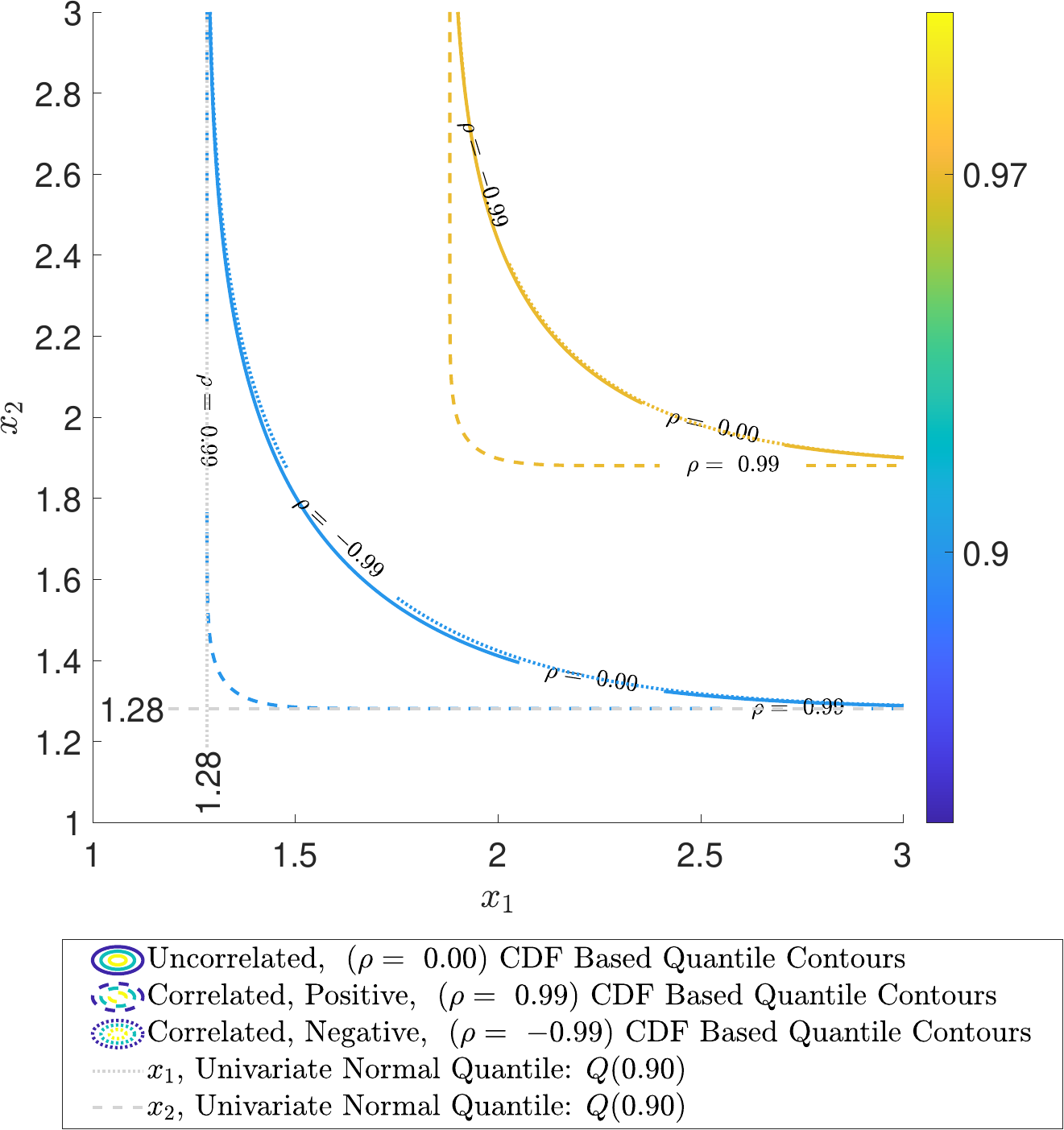}}\label{Plot:Bivariate_CDF_B}}\hspace{5pt}
\caption{The solid, dashed, and dotted contours represent correlation coefficients at, $\rho=\left\{0 , 0.99, -0.99  \right\}$ respectively. At low quantile probability values, there exists strong differences between the quantile contours for various correlation coefficients. At larger quantile probability values the differences between the quantiles for correlation coefficients $\rho=0$ and $\rho=-0.99$ become much harder to identify. At $\rho=0.99$ the quantile contours are still noticeably different compared to the quantile contours at $\rho=0$ and $\rho=-0.99$} \label{Plot:Bivariate_CDF}
\end{sidewaysfigure}

Figure \ref{Plot:Bivariate_PDF_CDF} builds off the example from Figure \ref{Plot:CorrelationContour} by showing some CDF based bivariate normal quantiles, $Q(\tau)$, from $\tau \in \left\{0.05, 0.1, \ldots, 0.90, 0.95 \right\}$, shown as a gradient of colored contours. Figure \ref{Plot:Bivariate_CDF} provides two different plots at three different quantile probability values, $\tau \in \left\{0.01, 0.90 , 0.97 \right\} $.  The solid, dashed, and dotted, contours represent correlation coefficients at, $\rho \in \left\{0 , 0.99, -0.99  \right\}$ respectively. At low quantile probability values, there exists strong differences between the quantile contours for various correlation coefficients, \ref{Plot:Bivariate_CDF_A}. At larger quantile probability values the differences between the quantiles from various correlation coefficients become harder to identify, \ref{Plot:Bivariate_CDF_B}. When $\tau = 0.90$, it is still easy to identify the case where $\rho=0.99$, as it is very close to the univariate normal quantiles of $Q(0.90)$. However, in the cases where $\rho=0$ and $\rho=-0.99$, it becomes much harder to distinguish between the two contours as there is a large amount of overlap, but they are different. The grey dashed lines in Figure \ref{Plot:Bivariate_CDF} represent the univariate normal quantiles, $Q(\tau) : \tau \in \left\{0.01,0.90\right\}$, which are the asymptotic bounds for the CDF based multivariate quantiles. Therefore, the CDF based multivariate quantiles are always strictly greater than or equal to each variate's univariate normal quantile. 

\afterpage{\clearpage}

\afterpage{\clearpage}

The figures below all utilize a bivariate normal distribution due to being easier to visualize and plot. For trivariate normal distributions, the quantiles are represented as iso-surfaces for constant quantile probability values. For multivariate distributions in $\mathbb{R}^q : q > 3$, the quantiles are hyper iso-surfaces.

\newpage
\section{Coverage for Multivariate Distributions}\label{coverage}
 Coverage, in the most general case, for multivariate distribution is taken from \eqref{EQ:TOL_UPPEReranceBeta} and is defined below,

\begin{gather}
\beta = \int \ldots \int_{\Omega}  \bigg( f(x_1, \ldots, x_q) \ dx_{1}, \ldots, dx_{q} \bigg) .   \label{EQ:Beta}
\end{gather}

 CDF based quantiles in a multivariate space are not intuitive for two main reasons. The first being ambiguity as there are many (infinitely) number of domains, $\Omega$, which satisfy \eqref{EQ:Beta}. Recall the bivariate case from \eqref{EQ:TOL_UPPEReranceBeta}, the domain is $\Omega = \left\{ \boldsymbol{x}=(x_1, x_2) \  | \  f(\boldsymbol{x},\bar{\boldsymbol{x}}, S) \le r \right\}$, which is an elliptical region. When the domain is $\Omega = \left\{ (x_1, x_2) \  | \ \boldsymbol{x} \le  \UT{T}_{\mathrm{simul}, \beta} \right\}$, the domain is a rectangular region in  $\mathbb{R}^2$ which can be seen as the upper right hand region starting at the intersection of the univariate normal quantiles in Figure \ref{Plot:EllipticalQuantile_vs_QuantileContour}, when the distribution parameter's are known.

\begin{figure}[h] 
\includegraphics[width= \textwidth]{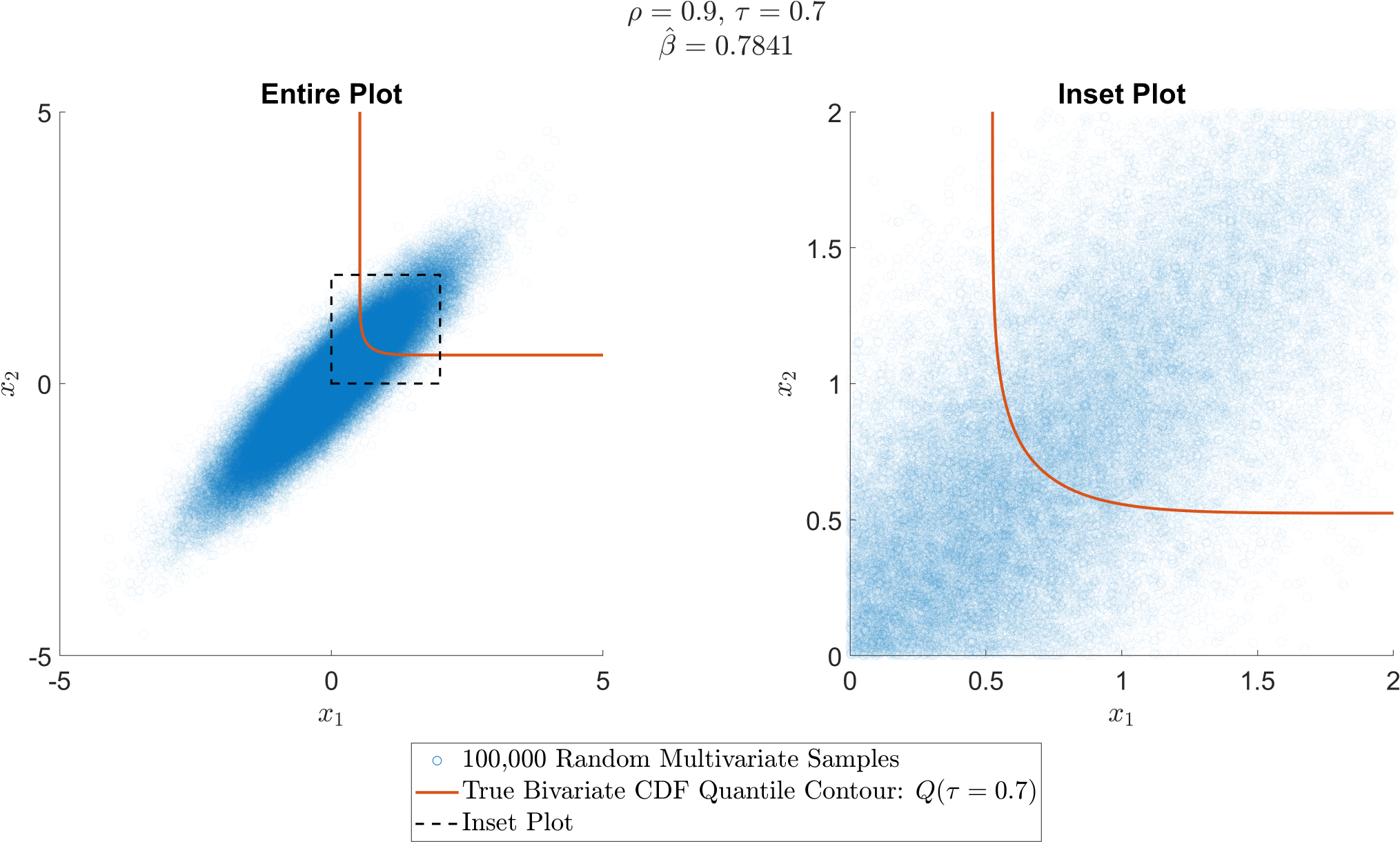}
\centering
\caption{Example of a standard normal bivariate distribution with a correlation of $\rho=0.9$ and a CDF quantile contour, $Q(\tau=0.7)$, and $\hat{\beta} = 0.7841$.}
\label{Plot:Numerical_Simulation_Example}
\end{figure}

The second reason is related to univariate statistics, where the $\tau^{\mathrm{th}}$ quantile probability also implies coverage, $\beta = \tau$, since $\beta = \int_{0}^{Q(\tau)} f(x) \,dx$. However, when the domain, $\Omega$, is changed from the CDF based quantile in \eqref{EQ:Beta} to $\Omega = \left\{ (x_1, x_2) \  | \ \boldsymbol{x} \le Q(\tau) \right\}$, the  equality now becomes an inequality,  $\beta \ge \tau$. Figure \ref{Plot:Numerical_Simulation_Example} provides an example for a standard normal bivariate distribution with a correlation of $\rho=0.9$ and a CDF quantile contour, $Q(\tau=0.7)$. Utilizing one hundred thousand random multivariate samples, the true value can be estimated accurately, $\hat{\beta} \approxeq \beta$. From Figure \ref{Plot:Numerical_Simulation_Example}, it is evidently clear that $\beta \ne \tau$, but $\beta \ge \tau$. The limiting case is when the correlation, $\rho$, approaches 1, then $\beta = \tau$, i.e.,$ \lim\limits_{\rho \to 1} \beta = \tau$. 

\end{appendix}

\bibliography{CI_MV_Reference}

\bigskip

\end{document}